\RequirePackage{fix-cm}
\documentclass[smallextended]{svjour3}
\smartqed
\usepackage{graphicx}
\usepackage{url}
\usepackage{natbib}
\usepackage{booktabs}
\usepackage{multirow}
\usepackage{amsmath}
\usepackage{amssymb}
\usepackage[table]{xcolor}
\definecolor{okgreen}{HTML}{1B7F3B}
\definecolor{nored}{HTML}{C0392B}
\definecolor{amber}{HTML}{D99400}
\definecolor{rowgray}{gray}{0.92}
\newcommand{\cyes}{\textcolor{okgreen}{$\boldsymbol{\checkmark}$}}
\newcommand{\cno}{\textcolor{nored}{$\boldsymbol{\times}$}}
\newcommand{\cpart}{\textcolor{amber}{$\boldsymbol{\sim}$}}
\makeatletter\def\makeheadbox{}\makeatother

\begin{document}

\title{Deep Agentic Search for Repository-Level Code Question Answering: An Empirical Study}
\titlerunning{Deep Agentic Search for Repo-Level Code QA}

\author{Amirkia Rafiei Oskooei \and Bora Ilci \and Alperen Kayim \and Mehmet Egemen Uzun \and Berat Can \and Kaan Emre Kara \and Ozan Orhan \and Mehmet S. Aktas}
\authorrunning{A. Rafiei Oskooei et al.}

\institute{A. Rafiei Oskooei \at
   Department of Computer Engineering, Yildiz Technical University, Istanbul, Turkey \\
   Intellica Business Intelligence, Istanbul, Turkey \\
   ORCID: 0009-0004-3490-550X \\
   \email{amirkia.oskooei@std.yildiz.edu.tr}, \email{amirkia.oskooei@intellica.net}
   \and
   B. Ilci (\email{bora.ilci@std.yildiz.edu.tr}), A. Kayim (\email{alperen.kayim@std.yildiz.edu.tr}), M. E. Uzun (\email{egemen.uzun@std.yildiz.edu.tr}), B. Can (\email{berat.karakas@std.yildiz.edu.tr}), K. E. Kara (\email{kaan.kara@std.yildiz.edu.tr}), O. Orhan (\email{ozan.orhan@std.yildiz.edu.tr}), M. S. Aktas (\email{aktas@yildiz.edu.tr}) \at
   Department of Computer Engineering, Yildiz Technical University, Istanbul, Turkey
}

\date{}

\maketitle

\begin{abstract}
   Code agents spend much of their effort simply locating the right code inside a repository. Two approaches dominate current practice. In Semantic Search, the agent retrieves code blocks from a vector index built from the repository in advance. In Deep Agentic Search (also known as grep-search by subagent), a planning agent delegates the exploration to a separate subagent that works in an isolated context window and returns only a condensed result. The second design, which is considered good context engineering practice, exists to protect the main agent from context pollution (also known as context rot), the loss of accuracy that occurs as unrelated material accumulates in the context window. Recent code agents (such as Claude Code, Codex, Antigravity, etc) have adopted it quickly, but there is little evidence on whether it produces better answers. We compare the two approaches on SWE-QA, a benchmark for repository-level code question answering. Semantic search answered 65.2\% of questions correctly against 46.2\% for deep agentic search, and it produced each correct answer at less than half the cost. To explain the gap, we then coded every failed run into a taxonomy of failure modes. The taxonomy shows that deep agentic search did not remove failures but introduced a new class of them: the single largest share of its failures, 41.8\%, occurred at the hand-off between the planner and its sub-agent, and these were usually silent, ending in a fluent and confident answer that was wrong. Deep agentic search addresses a real problem and is now the preferred design in many code agents. However, our results show that the protection it offers may not be free, and that for read-only questions over a repository that can be indexed, retrieval was the stronger and cheaper option.
\end{abstract}

\begin{figure}[!b]
   \centering
   \includegraphics[width=0.95\textwidth]{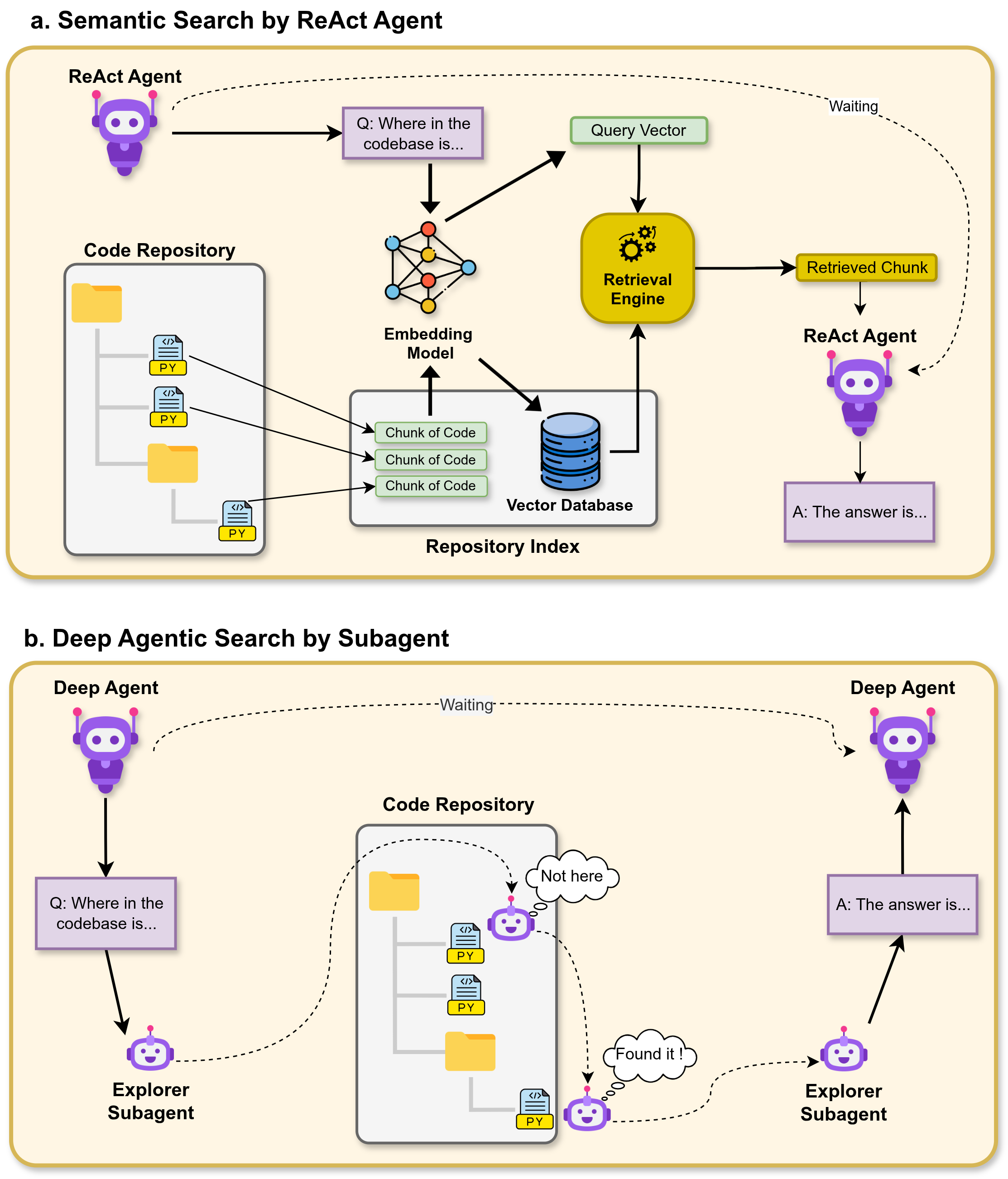}
   \caption{The two paradigms compared in this study. (a) Semantic search, realised by a ReAct agent, which retrieves pre-filtered chunks from a vector index built in advance. (b) Deep Agentic search, realised by a Deep Agent, which builds no index and delegates exploration to a sub-agent that works in its own context window and returns only a condensed result. The designs differ in what reaches the main context of the agent that answers.}
   \label{fig:hero}
\end{figure}

\section{Introduction}
\label{sec:intro}
Large language models have turned code agents into a primary interface for software development. Developers increasingly write, understand, debug, and maintain software through these agents, whether inside an integrated development environment or at the command line. One task recurs throughout this workflow, namely code question answering, which is the task of answering a natural language question about a codebase. A question can concern a single function, class, or snippet, or it can concern an entire repository. When the scope is a whole repository the task becomes repository-level code question answering, and it is the most demanding form of the task, because the information needed to answer a question is spread across many files and must first be located and then combined into a coherent answer. Both people and agents rely on this task. Developers ask questions about code in order to understand it before they extend, debug, or maintain it, and agents ask questions about code, either on behalf of a developer or for their own planning, in order to complete the tasks they are given.

Repository-level code question answering is costly for people and challenging for agents, and the difficulty grows with the size of the repository. The root of the difficulty is that an agent reasons with a language model that has a bounded context window and a fixed store of parametric knowledge, and neither of these holds a faithful and current copy of the specific repository at hand. Supplying more text is not a remedy on its own, because accuracy declines as unrelated material accumulates in the context window, an effect practitioners call context pollution, or context rot \citep{oskooei2026manyshot}. For each question, an agent must therefore decide which small part of a large and evolving codebase to place in front of the model. This problem of curating the right context, commonly called context engineering, is where the two paradigms we study diverge.

These two paradigms have become dominant in practice, and each is realised by a distinct agent architecture (Fig.~\ref{fig:hero}). The first is a ReAct agent that uses Semantic search, retrieving from a vector index built in advance from the repository. This is the established design behind the assistants embedded in modern editors, such as Cursor. The second is a Deep Agent that uses Deep Agentic search, also described by practitioners as grep search by an explorer sub-agent, exploring the repository on demand with terminal style commands and delegating the noisy part of that exploration to an isolated sub-agent that returns only a condensed result. Keeping the raw output of the search out of the main context in this way is a deliberate act of context engineering and is widely regarded as good practice. This is the more recent design behind assistants that operate at the command line, such as Claude Code, Gemini CLI, and Codex \citep{liu2026diveclaudecode,santos2026decoding}. The two paradigms embody a genuine design trade-off. Semantic search pays a one-time indexing cost and then serves compact, pre-filtered context, but it can fragment code and retrieve passages that are similar in wording yet irrelevant. Deep Agentic search pays no indexing cost and reads code in its natural structure, but it can flood the model with raw output that must then be filtered out again.

Repository exploration is a large part of an agent's work and consumes a substantial share of its tool calls and tokens, so the choice between these paradigms has direct consequences for accuracy, cost, and latency. Practitioners actively debate which paradigm is preferable, yet the debate rests largely on intuition and experience. To our knowledge the two paradigms have not been compared empirically for repository-level code question answering, and in particular no study has set an index-based retrieval agent against a Deep Agent that delegates repository exploration to an isolated sub-agent. This study provides that comparison. We do not propose a new method and we hold no prior preference for either paradigm. Our aim is not to recommend one paradigm over the other, but to measure where each one succeeds, where it fails, and at what cost, under the conditions we study.

We empirically compare these two paradigms, Semantic search and Deep Agentic search, on a standard benchmark for repository-level code question answering \citep{peng2026swe}. We organise the comparison around three research questions.
\begin{itemize}
   \item RQ1 (Accuracy): Which paradigm answers repository-level code questions more accurately, Semantic search or Deep Agentic search?
   \item RQ2 (Cost and efficiency): What does each paradigm cost, and is any difference in accuracy worth its cost?
   \item RQ3 (Behaviour and failure modes): How do the two paradigms behave, and how and why does each one fail?
\end{itemize}

This paper makes three contributions. First, it provides an empirical comparison of Semantic search and Deep Agentic search for repository-level code question answering across four language models on a standard benchmark. Second, it provides a cost-aware analysis that reports token consumption, monetary cost, latency, and the effective cost per correct answer, together with a view of the trade-off between accuracy and cost. Third, it provides a behavioural analysis of tool usage and trajectory length, and a taxonomy of failure modes derived from the recorded execution traces.
The remainder of the paper is organised as follows. Section~\ref{sec:background} provides background on the memory and context limits that motivate both paradigms and on how each paradigm works. Section~\ref{sec:related} discusses related work. Section~\ref{sec:design} describes the experimental design. Section~\ref{sec:results} reports the results. Section~\ref{sec:failure} presents the failure analysis. Section~\ref{sec:discussion} discusses the findings, Section~\ref{sec:ttv} states threats to validity, and Section~\ref{sec:conclusion} concludes.

\section{Background}
\label{sec:background}
This section provides the background needed to interpret the comparison. We first define repository-level code question answering and explain why context, rather than raw model capability, is the central constraint. We then describe the agent architectures and the search paradigms they use, and we close by positioning the two paradigms that our study compares.

\subsection{LLMs for Repository-Level Code Question Answering}
Code question answering is the task of answering a natural language question about source code. Its scope can be narrow, such as a question about a single function, class, or snippet, or wide, such as a question about an entire repository. Repository-level code question answering is the widest and most difficult form. A repository can contain hundreds of thousands of lines of code across many files, and the facts needed to answer a single question, for example how a class enforces validation or where a configuration value is consumed, are typically scattered across several of those files. Answering the question therefore has two parts. The agent must first locate the relevant code within the repository, and it must then understand and combine that code into a correct and coherent answer. The first part, locating the code, is what separates repository-level questions from narrower ones, and it is where much of an agent's effort is spent. Both developers and agents perform this task. Developers ask such questions to understand unfamiliar code before extending, debugging, or maintaining it, and agents ask them, on behalf of a developer or for their own planning, for the same reason.

An agent answers a question by prompting a language model, and the model draws on two kinds of memory. Parametric memory is the knowledge fixed in the model weights during training. Working memory is the context window, which is the text supplied to the model at inference time. Parametric memory alone is a poor basis for answering questions about a specific repository. Training a model on the full contents of large repositories is expensive, and even where it is possible, repositories change rapidly, so a model trained today becomes stale after the next commit and cannot be retrained after every change. Models also recall specific facts from their parameters less reliably than they read them from the context window, and answers produced from parametric memory alone are more prone to fabrication \citep{yang2025codesimpleqa}. To produce accurate and current answers, the relevant repository content must be placed in the context window.

The context window, however, is bounded, and a large repository does not fit within it and we need to be selective about what we put in it or summarize chunks of it \citep{oskooei2025hierarchical,rafiei2026natural}. Supplying more text is not a free remedy, because the quality of a model's answer tends to degrade as the input grows \citep{rando2025longcodebench,maharaj2026robustness,huang2026atlas}. As the context lengthens, the model must find the few relevant passages among many irrelevant ones, a difficulty often described as looking for a needle in a haystack. The amount of text is not the only factor. Model accuracy also falls as the proportion of irrelevant to relevant tokens rises, an effect described as context pollution, or context rot, in which distracting content crowds out the signal that the model needs. The effect is not confined to retrieved passages: in code translation, functional correctness peaks with a small number of in-context examples and then declines as many more are supplied, even when the context window is large enough to hold them \citep{oskooei2026manyshot}. Because the useful content is a small and shifting part of a large repository, the core problem for every paradigm is to select what enters the limited context window. This activity of curating the model's context is often called context engineering \citep{anthropic2025context,langchain2025context}, and it is the lens through which we interpret the paradigms that follow.

\subsection{Agent Architectures and Search Paradigms}
The first family of solutions is Semantic search, realised through retrieval-augmented generation \citep{lewis2020retrieval}. The repository is divided into chunks, an encoder model converts each chunk into a dense vector embedding, and the chunks and their embeddings are stored in a vector index. When a question arrives, it is embedded in the same space, and the chunks whose embeddings are most similar to the question, measured by a similarity such as cosine distance, are retrieved and placed in the context window for the model to reason over. This approach keeps the context compact, because only the most similar chunks are supplied, and it is inexpensive to keep current, because only the chunks affected by a change need to be re-embedded, rather than the whole model retrained. For these reasons it became a standard approach for repository-level code question answering \citep{gao2023rag}. In its agentic form the agent issues several retrieval queries in sequence rather than a single query, refining what it asks for as it reasons. This is commonly built as a ReAct agent \citep{yao2023react}, which follows a loop of reasoning, calling a retrieval tool, and observing the result until it can answer.

Semantic search also has characteristic weaknesses, and they follow from how the index is built. First, the quality of the answer is bounded by the quality of the retrieval, so when the retrieved chunks miss the relevant code the model can only produce an incomplete or incorrect answer. Second, code is hierarchical rather than linear, and dividing it into fixed-size chunks can split a function, separate a class from its methods, or remove the surrounding scope that a passage needs to be understood. An agent can request neighbouring chunks, but reassembling structure in this way is indirect. Third, following a dependency, such as moving from a class to the parent class it extends, calls for exact identifier matching, whereas retrieval reduces the request to an approximate similarity search, which adds imprecision to a task that requires precision. Fourth, in a large repository many code elements are worded similarly, so a query can match a large number of superficially similar passages and return results of low precision. Fifth, to compensate for this imprecision an agent may retrieve many candidate chunks, which raises token consumption and latency and can bury the correct passage among distractors, so that the model misses information that is present but lost in the middle of a long context.

The second family of solutions is Agentic search. Instead of consulting a pre-built index, the agent is given access to the file system and explores it on demand with terminal style commands such as \texttt{grep}, \texttt{find}, and \texttt{glob}. This mirrors how a developer navigates an unfamiliar codebase and removes the indexing step entirely. It also addresses several of the weaknesses of Semantic search. Because the agent reads files in their native form, it avoids the chunk boundaries that fragment code. Because it can match identifiers exactly, it can follow dependencies such as inheritance directly rather than through approximate search. Because it works on the live repository, it is always current and needs no ingestion or re-indexing. And because it targets literal identifiers and strings, it avoids returning a flood of merely similar passages in a large codebase.

Agentic search introduces a difficulty of its own. Exploration commands return large volumes of raw text, such as long file listings, directory trees, and unfiltered source, and this output accumulates in the agent's context window. Compared with the compact, pre-filtered passages of Semantic search, this raw material pollutes the context more quickly and pushes the agent toward the same degradation described above, while also raising token consumption and latency, because the agent must read through verbose output to find what matters.

\begin{figure}[htbp]
   \centering
   \includegraphics[width=0.80\textwidth]{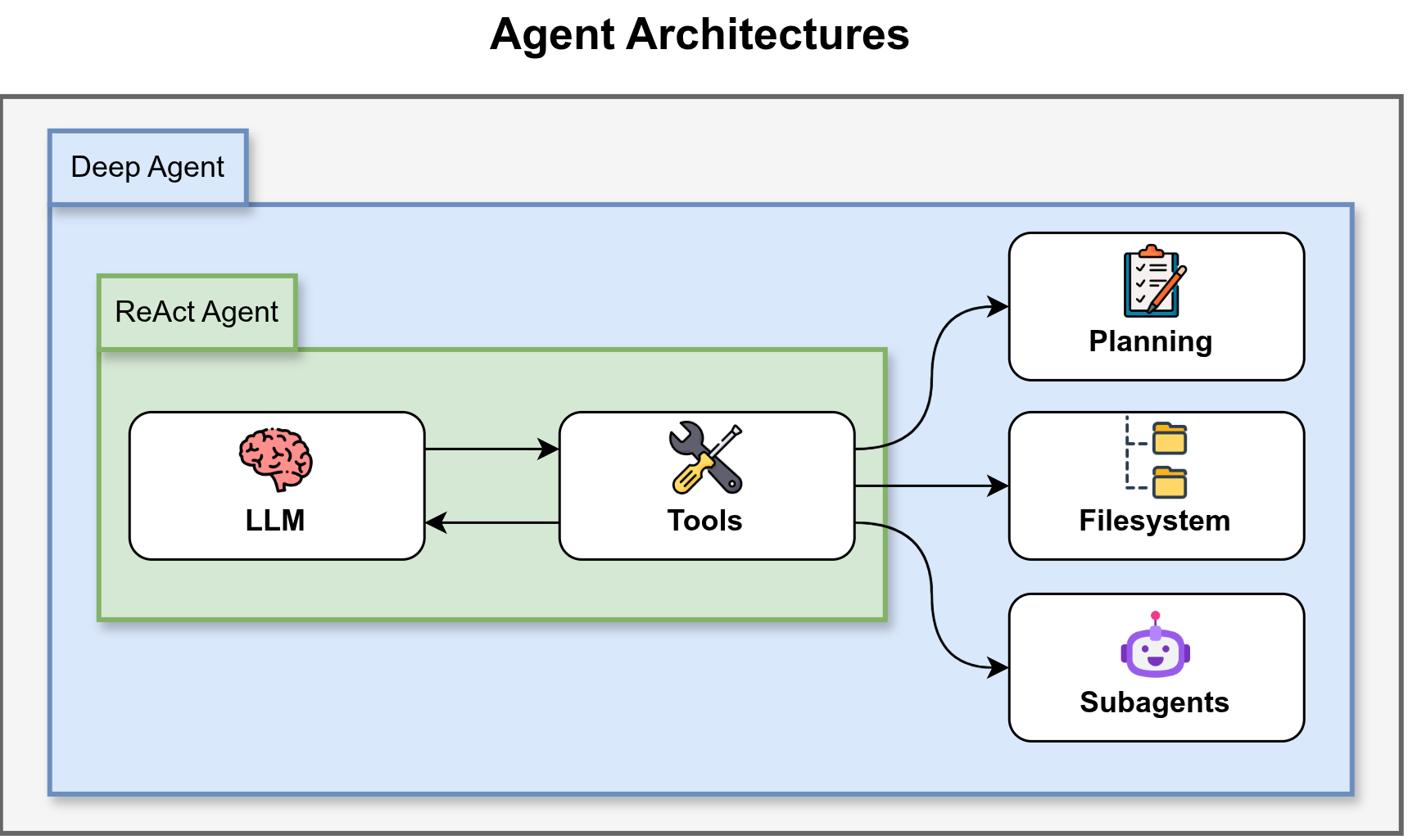}
   \caption{Agent architectures. A ReAct agent is a loop between a language model and a set of tools. A Deep Agent contains that same loop and extends the tool set with planning, a virtual filesystem, and the ability to delegate work to sub-agents.}
   \label{fig:arch}
\end{figure}

Deep Agentic search is a refinement that keeps the benefits of live exploration while containing this cost. Rather than have a single agent manage both high-level planning and low-level exploration, the agent is organised as a hierarchy. A main orchestrator plans the task and maintains a task list, and it delegates the noisy exploration loop to an isolated sub-agent. The sub-agent has its own separate context window in which it runs the exploration commands, absorbs the raw output, and filters it, and it returns only a condensed result to the orchestrator. The orchestrator's context therefore stays comparatively clean, and it can direct further exploration without accumulating the raw material itself. This delegation is itself a deliberate act of context engineering \citep{anthropic2025context}, and it is the design used by recent assistants that operate at the command line. The two architectures are nested rather than alternative (Fig.~\ref{fig:arch}). A Deep Agent runs the same reasoning and tool-calling loop as a ReAct agent, and extends it with planning, a virtual filesystem, and the ability to delegate work to a sub-agent.

\subsection{The Two Paradigms We Compare}
Our study compares the two designs that are most widely deployed today. The first is Semantic search, realised by a ReAct agent equipped with retrieval over a pre-built vector index. In this study the agent has exactly three tools, namely a repository structure lookup, a retrieval tool over the index, and a file reader, matching the action space of the original benchmark protocol \citep{peng2026swe}. It represents the indexing approach found in editor-based assistants such as Cursor and Antigravity. The second is Deep Agentic search, realised by a Deep Agent, a hierarchical architecture with an orchestrator that plans and delegates and a sub-agent that explores in isolation. It represents the approach found in command line assistants such as Claude Code, Gemini CLI, and Codex \citep{liu2026diveclaudecode,santos2026decoding}. Exploration-based designs that use terminal style tools without an index and without a sub-agent sit between the two and have been examined separately \citep{sen2026grep,zhang2026sweexplore}.

These two paradigms express opposite answers to the same context engineering question. Semantic search pays a one-time indexing cost and then serves compact, pre-filtered context, at the risk of fragmenting code and retrieving passages that are similar but irrelevant. Deep Agentic search pays no indexing cost and reads code in its native structure, at the risk of importing raw output that must be filtered, which the sub-agent delegation is designed to contain. The remainder of the paper measures how these opposing risks play out in accuracy, cost, and behaviour.

\section{Related Work}
\label{sec:related}
Prior work relevant to our study falls into three groups: benchmarks that define repository-level code question answering, systems that advance a single search paradigm, and the smaller set of studies that compare search paradigms directly. We review each group in turn and then state where our study differs. Table~\ref{tab:related} summarises the comparison along the dimensions that matter for our contribution.

\begin{table}[htbp]
   \centering
   \small
   \caption{Positioning against representative prior work. The columns are: Repo QA, repository-level code question answering; Paradigm comp., a direct comparison of semantic index-based search against agentic exploration-based search; Deep agent, the agentic side is a Deep Agent in which a planning orchestrator delegates exploration to an isolated sub-agent; Cost, a token, monetary, or latency analysis; Behaviour, an analysis of tool use or trajectory; Failure, a systematic failure-mode taxonomy. Here \cyes denotes that the dimension is addressed, \cpart that it is addressed in part, and \cno that it is not addressed.}
   \label{tab:related}
   \resizebox{\textwidth}{!}{%
      \begin{tabular}{lcccccc}
         \toprule
         Work                             & Repo QA & Paradigm comp. & Deep agent & Cost   & Behaviour & Failure \\
         \midrule
         \rowcolor{rowgray} SWE-QA        & \cyes   & \cpart         & \cno       & \cpart & \cno      & \cno    \\
         CoReQA                           & \cyes   & \cno           & \cno       & \cno   & \cno      & \cno    \\
         \rowcolor{rowgray} CodeRepoQA    & \cyes   & \cno           & \cno       & \cno   & \cno      & \cno    \\
         StackRepoQA                      & \cyes   & \cyes          & \cno       & \cno   & \cno      & \cno    \\
         \rowcolor{rowgray} SWE-Explore   & \cpart  & \cyes          & \cno       & \cpart & \cpart    & \cpart  \\
         Is Grep All You Need?            & \cno    & \cyes          & \cno       & \cno   & \cpart    & \cno    \\
         \rowcolor{rowgray} DeepRepoQA    & \cyes   & \cyes          & \cpart     & \cno   & \cno      & \cno    \\
         RepoSearch-R1                    & \cyes   & \cyes          & \cpart     & \cno   & \cno      & \cno    \\
         \rowcolor{rowgray} FastContext   & \cyes   & \cno           & \cyes      & \cyes  & \cno      & \cno    \\
         \midrule
         \rowcolor{rowgray} \textbf{Ours} & \cyes   & \cyes          & \cyes      & \cyes  & \cyes     & \cyes   \\
         \bottomrule
      \end{tabular}%
   }
\end{table}

\paragraph{Benchmarks for repository-level code question answering.}
A first line of work establishes the task and measures how well language models perform it. SWE-QA \citep{peng2026swe} provides open-ended questions over whole Python repositories and evaluates models under direct prompting, retrieval, and single-agent frameworks, and later work published under the same name extends the setting with multiple-choice questions \citep{elkoussy2026sweqa} and with a training recipe for open models \citep{cai2026sweqapro}. Throughout this paper, SWE-QA refers to the benchmark of \citet{peng2026swe}, which is the dataset we evaluate on. CoReQA \citep{chen2025coreqa} and CodeRepoQA \citep{hu2025coderepoqa} build repository-level question sets from issue threads and maintainer dialogues, and StackRepoQA \citep{alebachew2026beyond} draws repository questions from community answers, while LongCodeBench \citep{rando2025longcodebench} instead stresses very long context windows. These benchmarks fix the task and rank models, and where they vary the retrieval method they treat it as fixed infrastructure rather than as the object of study. The comparisons reported within the SWE-QA family in particular place retrieval against flat single agents, not against a hierarchical agent, and they report accuracy rather than a joint account of accuracy, cost, and failure. Adjacent benchmarks probe related capabilities, including code understanding that separates reasoning from documentation memorization \citep{zhang2026codeqabench}, code understanding and comparison \citep{anon2025codeinsightbench}, coding-agent evaluation beyond issue resolution \citep{raghavendra2026sweatlas}, developer-knowledge question answering \citep{zhang2026simpledevqa}, contextualized question answering for coding assistants \citep{mohammed2026rubberduck}, multi-domain coding question answering \citep{rahman2025refactorcoderqa}, repository documentation quality via question answering \citep{wang2026swdbench}, and code information retrieval and search \citep{li2025coir,gong2025cosqa}.

\paragraph{Systems that advance a single search paradigm.}
A second line proposes and improves one paradigm at a time. On the semantic side, retrieval-augmented generation, introduced for knowledge-intensive natural language tasks \citep{lewis2020retrieval} and since surveyed in depth \citep{gao2023rag}, including iterative multi-step variants \citep{wang2025corag}, is the basis of index-based code assistants, and code-specific systems such as CoRAC \citep{choi2025corac} and RepoWise \citep{kocadere2026repowise} adapt retrieval to code question answering. On the agentic side, the ReAct design \citep{yao2023react} established the reasoning and acting loop that our Semantic search agent instantiates, and more recent systems explore the repository dynamically rather than through an index: DeepRepoQA \citep{peng2026deeprepoqa} and RepoSearch-R1 \citep{li2025reposearch} navigate the repository with tree-search agents, and FastContext \citep{zhang2026fastcontext} and ContextPilot \citep{gao2026contextpilot} pair a lightweight explorer with a separate generator that consumes its condensed output. A related group of efficiency-oriented methods, including FastCode \citep{li2026fastcode}, SWE-Pruner \citep{wang2026swepruner}, and LongCodeZip \citep{shi2025longcodezip}, reduces the token cost of repository context. Graph-based retrieval agents add code-graph structure to exploration \citep{shah2025ranger,hu2026larger,usai2026logiclens}, general reasoning agents scale tool use \citep{li2026deepagent}, structured reasoning improves the reliability of code agents \citep{ugare2026agentic}, and further context-pruning and compression methods trim retrieved code before it reaches the model, including approaches that change its representation \citep{wang2026pruning,feng2026compression,shi2026codeocr}. Earlier code question answering operated at the level of single functions or snippets, using fine-tuned models \citep{andryushchenko2024codeqa,lopes2025t5} and retrieval-augmented pipelines \citep{ahmed2024codeqa}, with benchmarks measuring comprehension at that scale \citep{havare2025comprehension}. Each of these is a method contribution that advances a single paradigm, typically with one model, and does not set one paradigm against the other under matched conditions.

\paragraph{Direct comparisons of search paradigms.}
A third and smaller line compares paradigms head to head, and it is closest to our study. The work asking whether grep is all that is needed \citep{sen2026grep} finds that lexical exploration can match or exceed vector retrieval inside agent harnesses, but it studies conversational memory rather than code repositories. SWE-Explore \citep{zhang2026sweexplore} compares classical retrieval against agentic exploration at repository scale and finds exploration stronger, but it measures line-level localization for downstream patching rather than the correctness of an answer. StackRepoQA \citep{alebachew2026beyond} compares retrieval, agentic, and graph augmentation for repository question answering, but its agentic condition is a flat single agent. In all three, the agentic side is a single agent rather than a Deep Agent, and none of them pairs the comparison with a cost, a behaviour, and a failure analysis.

\paragraph{Positioning.}
Our study occupies the cell that these three groups leave empty (Table~\ref{tab:related}). To our knowledge, it is the first to set a Deep Agent that uses Deep Agentic search, a planning orchestrator that delegates repository exploration to an isolated sub-agent, against a ReAct agent that uses Semantic search, on repository-level code question answering. This delegated-explorer design mirrors how command-line assistants offload exploration to a smaller explorer agent, and it is distinct from the flat single agents used in the prior comparisons above. We further accompany the accuracy comparison with three analyses that prior work reports at most in part: a cost and efficiency analysis in tokens and dollars that includes the cost per correct answer, an analysis of agent behaviour through tool use and trajectory length, and a systematic taxonomy of failure modes. The combination of the Deep Agentic search paradigm with these three analyses, across four language models on a shared benchmark, is what distinguishes our study from the work reviewed above.

\section{Experimental Design}
\label{sec:design}
This empirical study was conducted as a collaborative research effort between Intellica Business Intelligence and researchers at Yildiz Technical University. We compare two agent architectures, a ReAct agent that uses Semantic search and a Deep Agent that uses Deep Agentic search. Each question is answered in a fresh conversation thread, so that no question can contaminate another, and only the repository targeted by the question is available to the agent. All execution traces are recorded, including tool calls, planner steps, sub-agent actions, and token accounting. Each research question is served by a specific part of this design: the graded answer quality and the paired statistical tests for RQ1, the token and cost measurements for RQ2, and the recorded traces, from which we derive tool usage, trajectory length, and a taxonomy of failure modes, for RQ3.

\subsection{Agent Implementation}
\label{sec:agentimpl}
We compare two agent architectures. The first is a ReAct agent that uses the Semantic search paradigm. It is the established reasoning and acting design, and we use the ReAct agent harness from LangChain to implement SWE-QA Agent, restricting it to the three tools of the original benchmark protocol: a repository structure lookup, a retrieval tool over a vector index, and a file reader. It has no shell access, no filesystem writing, and no planning tools. The vector index is built once per repository by splitting each file into chunks, embedding the chunks with an encoder model, and storing them for similarity retrieval at query time. The second is a Deep Agent that uses the Deep Agentic search by subagent paradigm. It is a more recent hierarchical design, and we use the Deep Agent harness from the Deep Agents framework, also from LangChain, with its default toolset, which includes planning, a virtual filesystem, and sub-agent delegation. The retrieval configuration for the ReAct agent, namely chunk size, overlap, index, similarity threshold, and embedding model, and the full tool contract of the Deep Agent, are given in Appendices~\ref{app:prompts} and~\ref{app:rag}. Figure~\ref{fig:actionspace} sets the two action spaces side by side.

The two action spaces are not identical, and they cannot be. The difference in tools is not a confound that sits alongside the paradigms; it is what constitutes them. A ReAct agent that was given planning, a task list, and a delegable sub-agent would be a Deep Agent, and a Deep Agent stripped of those and handed a vector index would be a ReAct agent. Equalising the tool sets would therefore not produce a cleaner comparison, it would collapse the two conditions into one. What we hold fixed instead is everything that is not constitutive of the paradigms: the same fifteen repositories, the same 720 questions, the same four models, the same judge, and the same rubric, with each model compared against itself across the two architectures so that the contrast is paired within a model rather than across models.

We also fixed the framework. Both agents are built on LangChain, using its ReAct agent harness for the first and its Deep Agents harness for the second, rather than pairing a research implementation against a commercial assistant. LangChain is open source, model-agnostic, and not tied to a single vendor's models or editor, and it is at the time of writing the most widely used framework for building agents, which makes it a neutral common substrate and one whose defaults a reader can inspect. Using each harness with its default tool contract, rather than a tuned variant of our own, also keeps the comparison closer to what a practitioner adopting either paradigm would actually run. Both harnesses remain choices, and Section~\ref{sec:ttv} states what follows from that.

\begin{figure}[htbp]
   \centering
   \includegraphics[width=0.92\textwidth]{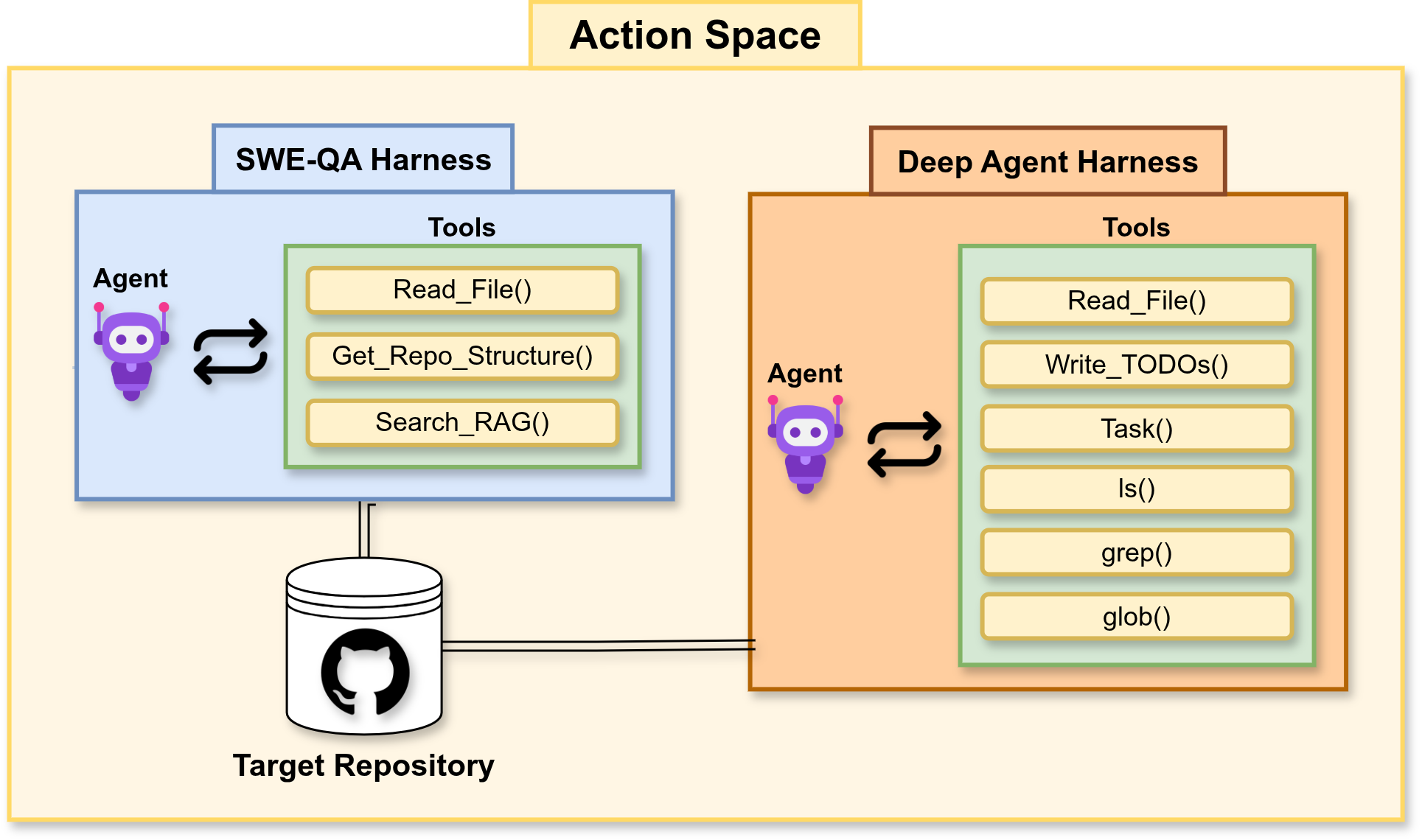}
   \caption{The action spaces of the two agents. Both act on the same target repository, and they differ in what they are allowed to do to it. The ReAct agent is held to the three tools of the benchmark protocol, namely a file reader, a repository structure lookup, and retrieval over the vector index. The Deep Agent has the default tool set of its harness, which replaces retrieval with terminal style exploration and adds planning and delegation to a sub-agent.}
   \label{fig:actionspace}
\end{figure}

\subsection{Models}
We run both architectures with four language models. Three come from a single widely used family, namely Gemini 2.5 Flash, Gemini 2.5 Pro, and Gemini 3 Flash. Sampling within one family lets us observe the paradigm effect across different capability tiers and across model generations while holding the provider fixed. The fourth, Qwen3-235B, is a strong open-weight model from a different vendor with a different architecture, and it serves as a cross-vendor check that any effect we observe is not specific to a single provider. Within the Deep Agent, the orchestrator and the sub-agent use the same model. This yields eight conditions, four models crossed with the two architectures. Model access, temperature, and provider details are given in Appendix~\ref{app:env}. We derive monetary cost from measured token counts using a single price sheet, expressed in USD per one million tokens as input and output: Gemini 2.5 Flash at 0.30 and 2.50, Gemini 2.5 Pro at 1.25 and 10.00, Gemini 3 Flash at 0.50 and 3.00, and Qwen3-235B at 0.22 and 0.88. Reasoning tokens are billed at the output rate. Because monetary cost depends on prices that change over time, we also report the underlying token counts, which are price-independent.

\subsection{Dataset}
We use SWE-QA \citep{peng2026swe}, a benchmark for repository-level code question answering over widely used open-source Python projects. Our evaluation covers 15 Python repositories with 48 questions each, for a total of 720 questions per condition. Twelve of these repositories are drawn from SWE-bench and three from SWE-bench-Live, which is built from more recent issues. The repositories span a wide range of sizes, from small libraries of about 13 thousand lines of Python to large projects of more than 800 thousand lines. The questions fall into four interrogative categories, namely What, Why, Where, and How, in roughly equal proportion. We use the dataset and its rubric as published and do not modify either.

\subsection{Evaluation}
Answers are graded by an LLM judge, Claude Sonnet 4.6, which comes from a different provider and a different model family than every answering model, so no condition is graded by its own model or by a relative of it. The judge scores each answer on the five SWE-QA dimensions, namely correctness, completeness, relevance, clarity, and reasoning. The five dimension scores are combined into a total on a 0 to 100 scale, and each answer receives a three-class verdict, namely Pass, Partial, or Fail, using the benchmark thresholds of Pass at 70 or above, Partial from 50 up to 70, and Fail below 50. The verdict bands are applied consistently across all conditions. The full judge prompt and the score aggregation are given in Appendix~\ref{app:judge}.

Because the two architectures answer the same questions, we compare them with paired tests within each model. We test the Pass rate difference with the exact McNemar test and the score difference with the Wilcoxon signed-rank test. We report Cliff's delta as a non-parametric effect size for the score difference, with a 95\% bootstrap confidence interval, and the paired standardised mean difference $d_z$. To control the family-wise error rate across the four models, we report Benjamini and Hochberg false discovery rate corrected values.

To check that the automatic judge is a sound proxy for human judgement, we validated it against an independent human panel on a stratified sample. Three external software engineers, each with at least two years of professional experience and none a co-author of this paper, rated a sample of 800 answers drawn evenly across the eight conditions, that is 100 answers per condition, from the same fifteen repositories. The raters were blind to the condition, the architecture, and the model behind each answer. Each rater independently assigned a Pass, Partial, or Fail verdict, using the same three outcome categories as the main study, together with a holistic score on a 0 to 100 scale. We then compared the human panel against the LLM judge.

Table~\ref{tab:judgevalidity} reports the agreement. The internal agreement of the panel was almost perfect on the verdict, with a Fleiss' kappa of 0.81. Agreement between the human majority and the LLM judge was substantial on the verdict, with a Cohen's kappa of 0.74, and strong on the numeric score, with a Pearson correlation of 0.83 and a mean absolute difference of 5.2 points on the 0 to 100 scale. Agreement was slightly lower on Deep answers than on Semantic answers, for example a kappa of 0.69 against 0.78 on the verdict, which is consistent with the greater length and complexity of Deep answers. On the scale of \citet{landis1977}, these values fall in the substantial to almost perfect range, which supports the use of the automatic judge as a reliable proxy for human judgement across both paradigms.

\begin{table}[htbp]
   \centering
   \small
   \caption{Judge validity: agreement between the independent human panel and the LLM judge on a stratified sample of 800 answers, 100 per condition. The upper block reports the three-class verdict and the lower block reports the harmonized 0 to 100 score. The panel-internal row uses Fleiss' kappa and the human against judge rows use Cohen's kappa. MAD is the mean absolute difference in score points.}
   \label{tab:judgevalidity}
   \begin{tabular}{lrrl}
      \toprule
      \multicolumn{4}{l}{Categorical verdict (Pass, Partial, Fail)}                             \\
      \midrule
      Comparison                                   & N   & Exact agr. (\%) & $\kappa$ [95\% CI] \\
      \midrule
      Human panel (Fleiss' $\kappa$, three raters) & 800 & 88.5            & 0.81 [0.78, 0.84]  \\
      Human majority vs LLM judge, pooled          & 800 & 84.0            & 0.74 [0.70, 0.78]  \\
      \quad Semantic conditions only               & 400 & 87.0            & 0.78 [0.72, 0.84]  \\
      \quad Deep conditions only                   & 400 & 81.0            & 0.69 [0.63, 0.75]  \\
      \bottomrule
   \end{tabular}

   \vspace{0.7em}

   \begin{tabular}{lrllr}
      \toprule
      \multicolumn{5}{l}{Numeric score (harmonized 0 to 100)}                               \\
      \midrule
      Comparison                      & N   & Pearson $r$ [95\% CI] & Spearman $\rho$ & MAD \\
      \midrule
      Human mean vs LLM judge, pooled & 800 & 0.83 [0.81, 0.85]     & 0.81            & 5.2 \\
      \quad Semantic conditions only  & 400 & 0.86 [0.83, 0.88]     & 0.85            & 4.1 \\
      \quad Deep conditions only      & 400 & 0.79 [0.75, 0.82]     & 0.76            & 6.8 \\
      \bottomrule
   \end{tabular}
\end{table}

\section{Results}
\label{sec:results}
This section reports the results on accuracy, cost, and agent behaviour; a dedicated failure analysis follows in Section~\ref{sec:failure}. Table~\ref{tab:master} summarises accuracy, efficiency, and cost for all eight conditions. Throughout, we refer to the Semantic search paradigm as Semantic and to the Deep Agentic search paradigm as Deep.

\subsection{Accuracy and answer quality}
\label{sec:rq1}
\begin{figure}[htbp]
   \centering
   \includegraphics[width=0.95\textwidth]{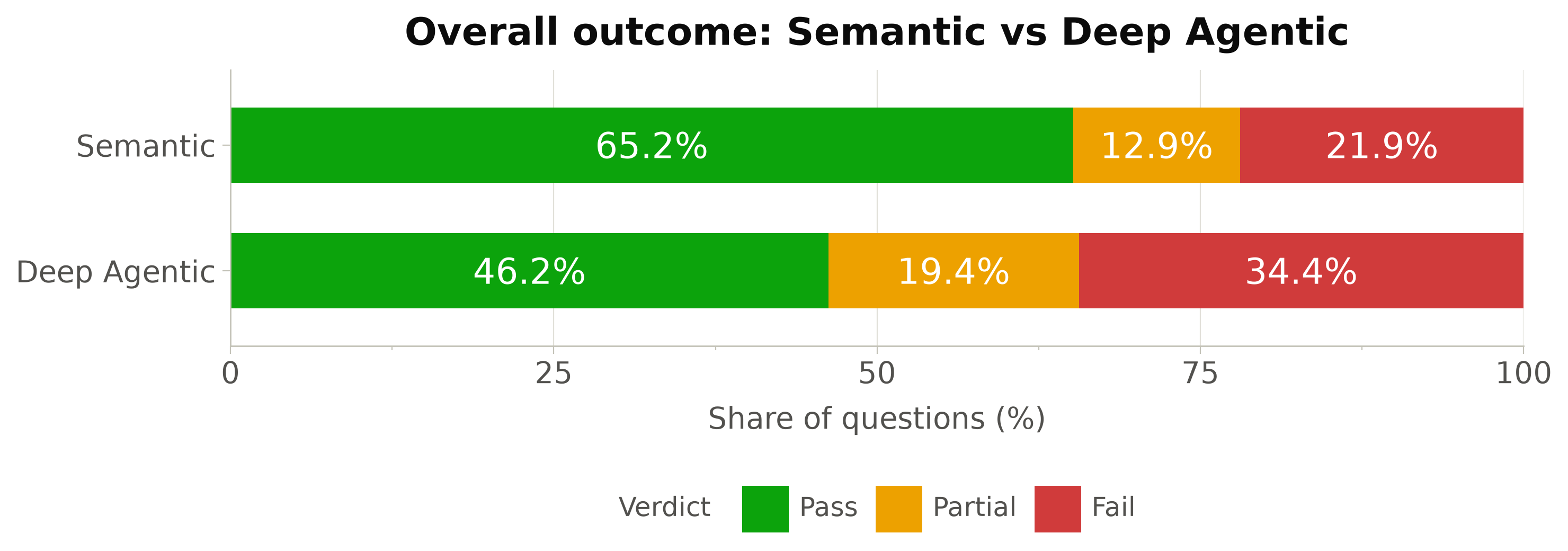}
   \caption{Overall outcome pooled across the four models. Semantic answers a larger share of questions at the Pass level and a smaller share at the Fail level than Deep.}
   \label{fig:overall}
\end{figure}

\begin{figure}[htbp]
   \centering
   \includegraphics[width=0.9\textwidth]{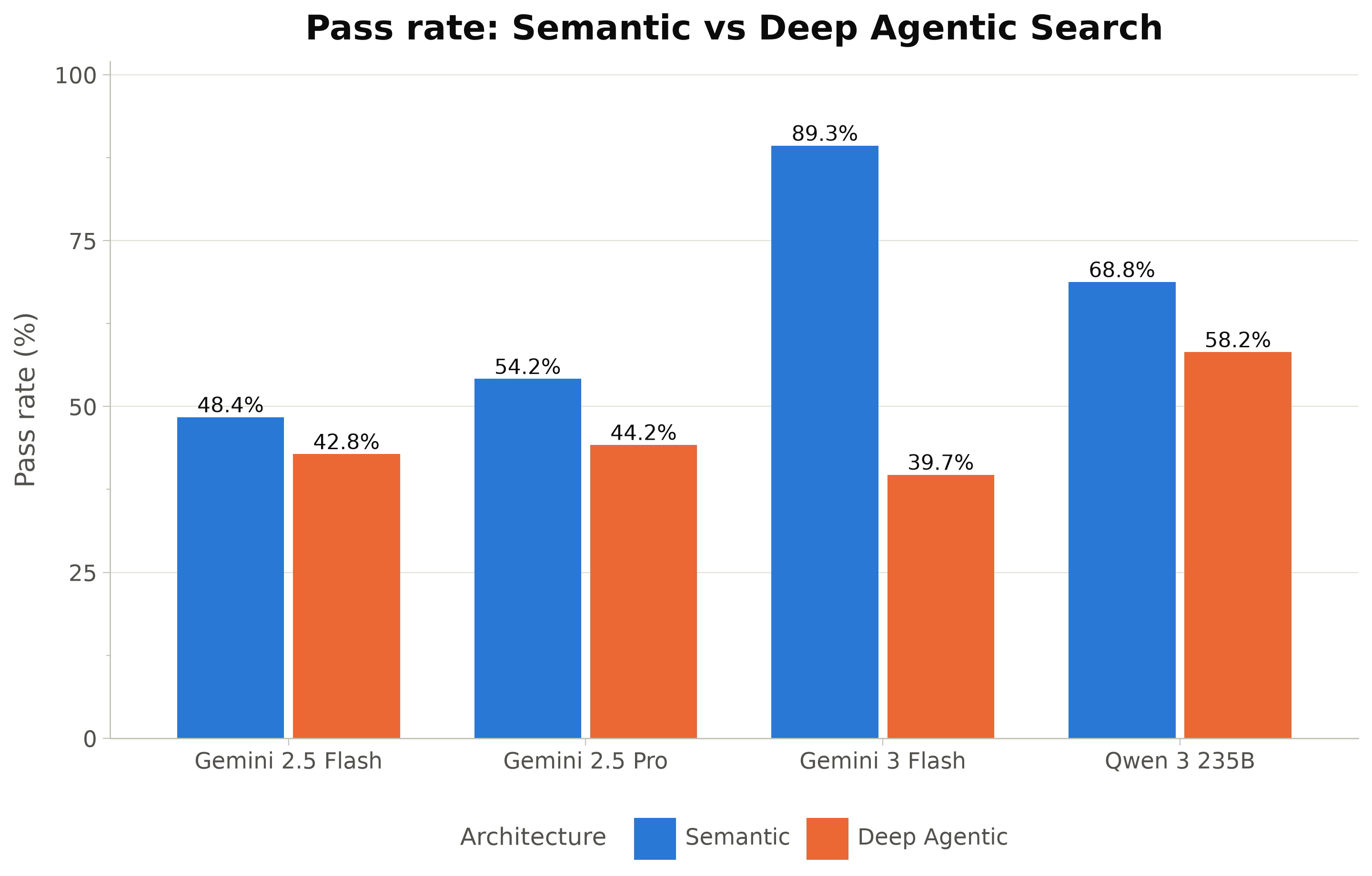}
   \caption{Pass rate of Semantic and Deep for each model. Semantic is higher for every model, with the largest gap on Gemini 3 Flash.}
   \label{fig:accuracy}
\end{figure}

Our first research question concerns accuracy. Pooled across the four models, Semantic answered a larger share of questions correctly than Deep. Semantic reached a Pass rate of 65.2\% against 46.2\% for Deep, and a Fail rate of 21.9\% against 34.4\% for Deep (Fig.~\ref{fig:overall}). The ordering held for every individual model (Fig.~\ref{fig:accuracy}, Table~\ref{tab:master}): Semantic obtained a higher Pass rate and a higher mean score than Deep in all four cases.

\begin{table}[t]
\centering
\footnotesize
\caption{Master comparison per condition: accuracy (Pass/Partial/Fail, mean judge score /100), efficiency (tokens, tool calls, median latency), and cost. The token columns report prompt and completion tokens only. Reasoning tokens are billed at the completion rate and are included in the cost, so the cost column is not the product of the two token columns and the list prices.}
\label{tab:master}
\resizebox{\textwidth}{!}{%
\begin{tabular}{llrrrrrrrrrr}
\toprule
Model & Search & Pass \% & Partial \% & Fail \% & Score & In-tok & Out-tok & Tools & Lat (s) & Mean \$/q & \$/Pass \\
\midrule
\multirow{2}{*}{Gemini 2.5 Flash} & Semantic & 48.4 & 15.7 & 35.9 & 61.6 & 303k & 1k & 9.7 & 329 & 0.1029 & 0.213 \\
 & Deep Agentic & 42.8 & 14.0 & 43.1 & 53.7 & 524k & 4k & 20.6 & 106 & 0.1868 & 0.436 \\
\addlinespace
\multirow{2}{*}{Gemini 2.5 Pro} & Semantic & 54.2 & 14.7 & 31.1 & 63.9 & 404k & 1k & 11.4 & 400 & 0.5675 & 1.048 \\
 & Deep Agentic & 44.2 & 17.1 & 38.7 & 57.2 & 466k & 4k & 15.5 & 152 & 0.6749 & 1.526 \\
\addlinespace
\multirow{2}{*}{Gemini 3 Flash} & Semantic & 89.3 & 5.4 & 5.3 & 89.2 & 254k & 3k & 8.6 & 101 & 0.1601 & 0.179 \\
 & Deep Agentic & 39.7 & 27.2 & 33.1 & 58.3 & 577k & 3k & 44.6 & 262 & 0.3407 & 0.858 \\
\addlinespace
\multirow{2}{*}{Qwen 3 235B} & Semantic & 68.8 & 15.8 & 15.4 & 77.0 & 34k & 1k & 3.5 & 42 & 0.0085 & 0.012 \\
 & Deep Agentic & 58.2 & 19.0 & 22.8 & 69.3 & 761k & 2k & 40.1 & 105 & 0.1693 & 0.291 \\
\bottomrule
\end{tabular}%
}
\end{table}

\begin{figure}[htbp]
   \centering
   \includegraphics[width=\textwidth]{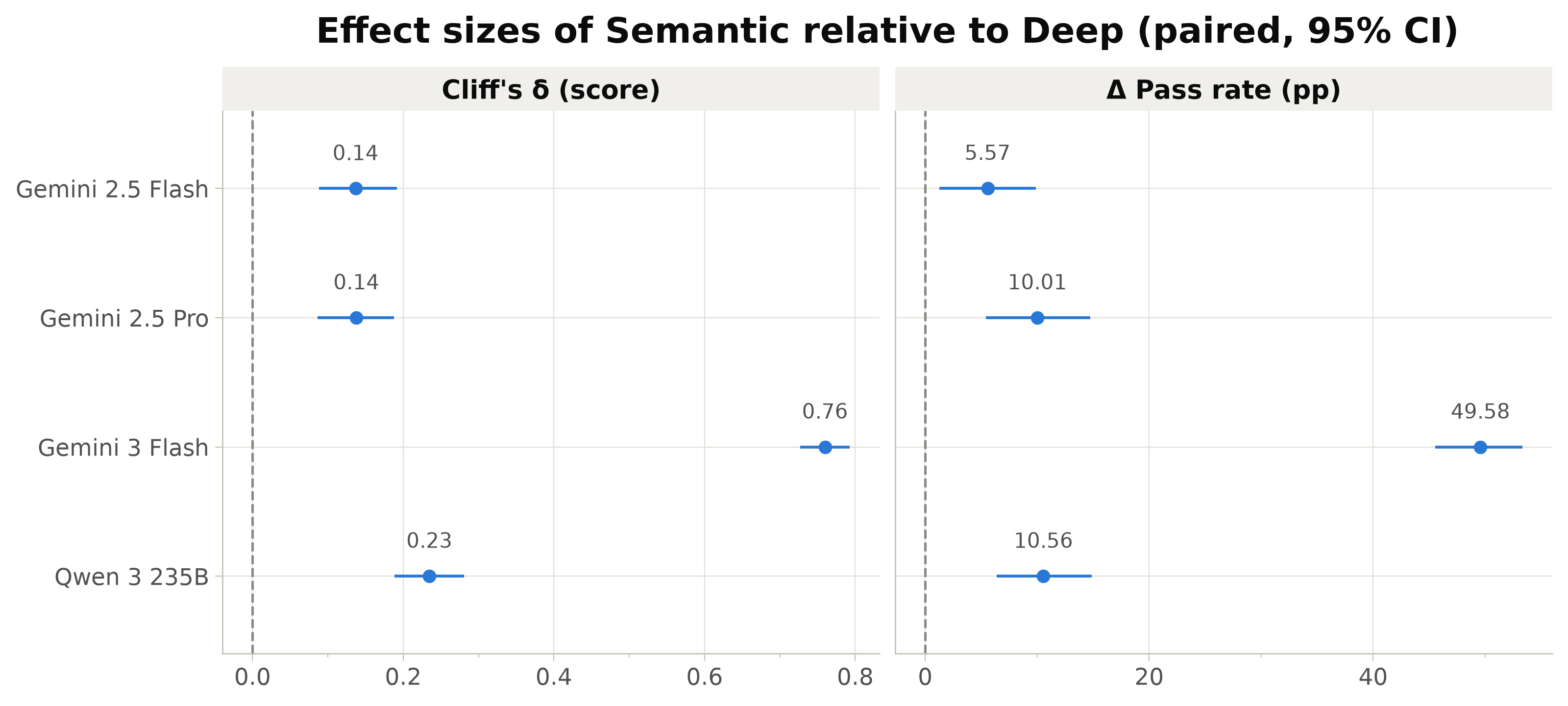}
   \caption{Paired effect sizes of Semantic relative to Deep with 95\% confidence intervals. Positive values favour Semantic. Every result remains significant after false discovery rate correction.}
   \label{fig:forest}
\end{figure}

The size of the advantage varied considerably by model. The paired within-model comparison (Table~\ref{tab:effects}, Fig.~\ref{fig:forest}) shows a Pass rate difference in favour of Semantic of 5.6 percentage points for Gemini 2.5 Flash, 10.0 for Gemini 2.5 Pro, 10.6 for Qwen3-235B, and 49.6 for Gemini 3 Flash. All four McNemar tests and all four Wilcoxon tests remained significant after false discovery rate correction. The effect size on the score, measured by Cliff's delta, was about 0.14 for the two Gemini 2.5 models, 0.24 for Qwen3-235B, and 0.76 for Gemini 3 Flash. We therefore describe the accuracy result as consistent in direction across all four models, modest in magnitude for the Gemini 2.5 pair, and large only for Gemini 3 Flash.

\begin{figure}[htbp]
   \centering
   \includegraphics[width=0.92\textwidth]{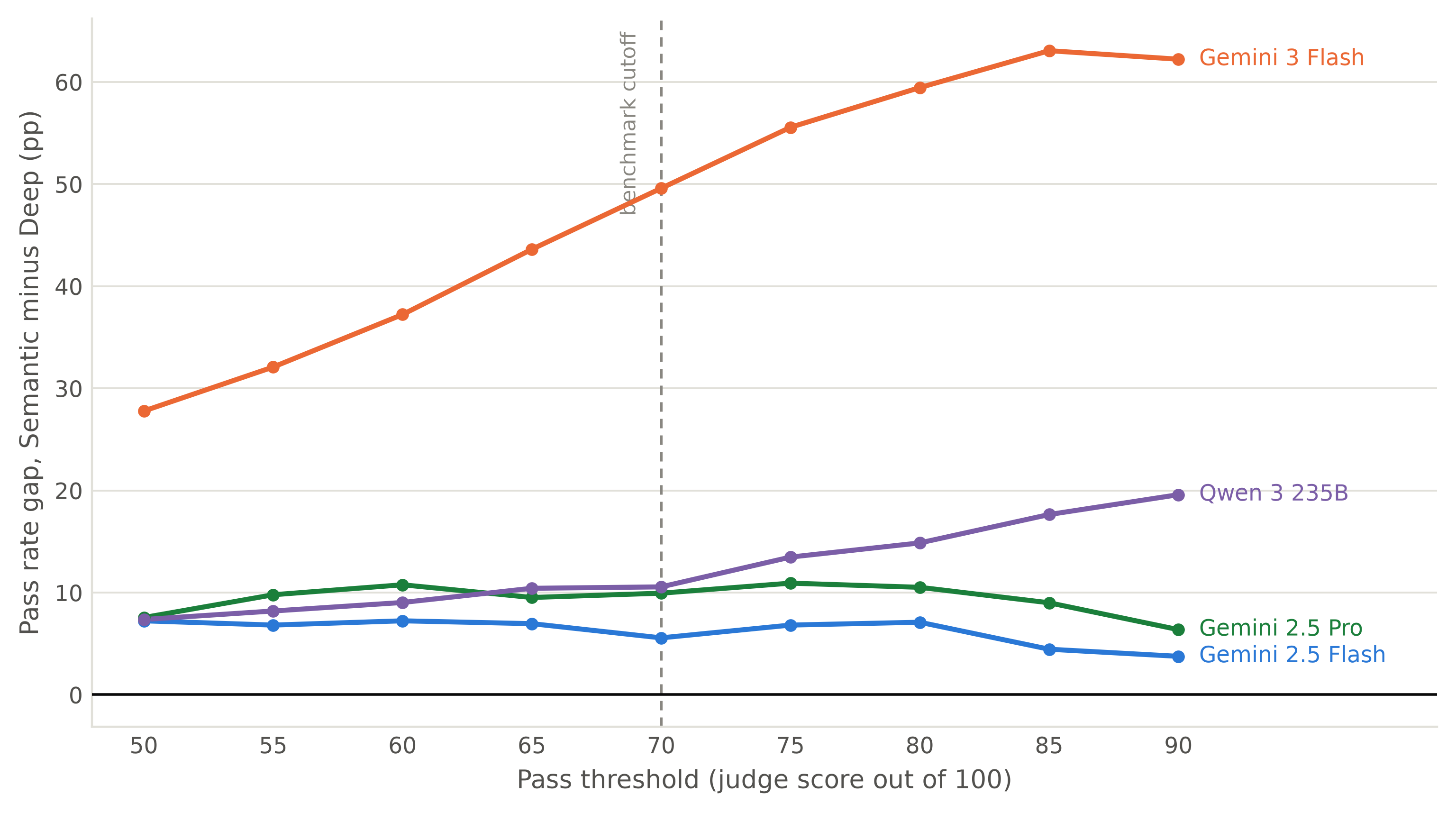}
   \caption{Sensitivity of the accuracy comparison to the Pass threshold. Each line is the Pass rate of Semantic minus the Pass rate of Deep for one model, as the cutoff is swept from 50 to 90 on the 0 to 100 judge scale. The dashed line marks the benchmark cutoff of 70. The magnitude of the gap depends on the cutoff, but every line stays above zero, so the ordering does not.}
   \label{fig:threshold}
\end{figure}

Because the Pass rate is defined by a fixed cutoff, part of a difference expressed in percentage points reflects where the score distributions sit relative to that cutoff rather than the size of the underlying difference. This is most visible for Gemini 3 Flash, where a score difference of 30.9 points corresponds to a Pass rate difference of 49.6 percentage points, because a large share of its answers falls close to the threshold. We therefore repeated the comparison across the full range of plausible cutoffs (Fig.~\ref{fig:threshold}). Sweeping the Pass threshold from 50 to 90 changes the magnitude of the gap considerably, from 27.8 to 63.1 percentage points for Gemini 3 Flash and from 7.4 to 19.6 for Qwen3-235B, but it does not change the ordering. Semantic exceeded Deep at every threshold for every model, and the smallest gap anywhere on the grid was 3.8 percentage points. The threshold-free measures reported above, namely the Wilcoxon test on the scores and Cliff's delta, do not depend on this choice, and we treat them as the primary evidence for the accuracy result.

\begin{table}[htbp]
   \centering
   \small
   \caption{Paired within-model comparison of Semantic against Deep. Positive values indicate that Semantic is higher. The Pass rate difference is in percentage points and the score difference is on a 0 to 100 scale, each with a 95\% bootstrap confidence interval. The McNemar value is corrected for false discovery rate across the four models.}
   \label{tab:effects}
   \resizebox{\textwidth}{!}{%
      \begin{tabular}{lrlclcr}
         \toprule
         Model            & N   & Pass diff (pp)    & McNemar $q$         & Score diff        & Cliff's $\delta$     & $d_z$ \\
         \midrule
         Gemini 2.5 Flash & 720 & 5.6 [1.3, 9.8]    & $1.2\times10^{-2}$  & 8.0 [5.3, 10.6]   & 0.137 [0.090, 0.186] & 0.220 \\
         Gemini 2.5 Pro   & 720 & 10.0 [5.4, 14.6]  & $4.5\times10^{-5}$  & 6.7 [4.2, 9.4]    & 0.138 [0.089, 0.188] & 0.190 \\
         Gemini 3 Flash   & 720 & 49.6 [45.6, 53.3] & $3.4\times10^{-93}$ & 30.9 [29.0, 33.0] & 0.760 [0.727, 0.793] & 1.125 \\
         Qwen3-235B       & 720 & 10.6 [6.4, 14.9]  & $2.9\times10^{-6}$  & 7.7 [5.9, 9.7]    & 0.235 [0.191, 0.283] & 0.296 \\
         \bottomrule
      \end{tabular}%
   }
\end{table}

\begin{figure}[htbp]
   \centering
   \includegraphics[width=0.88\textwidth]{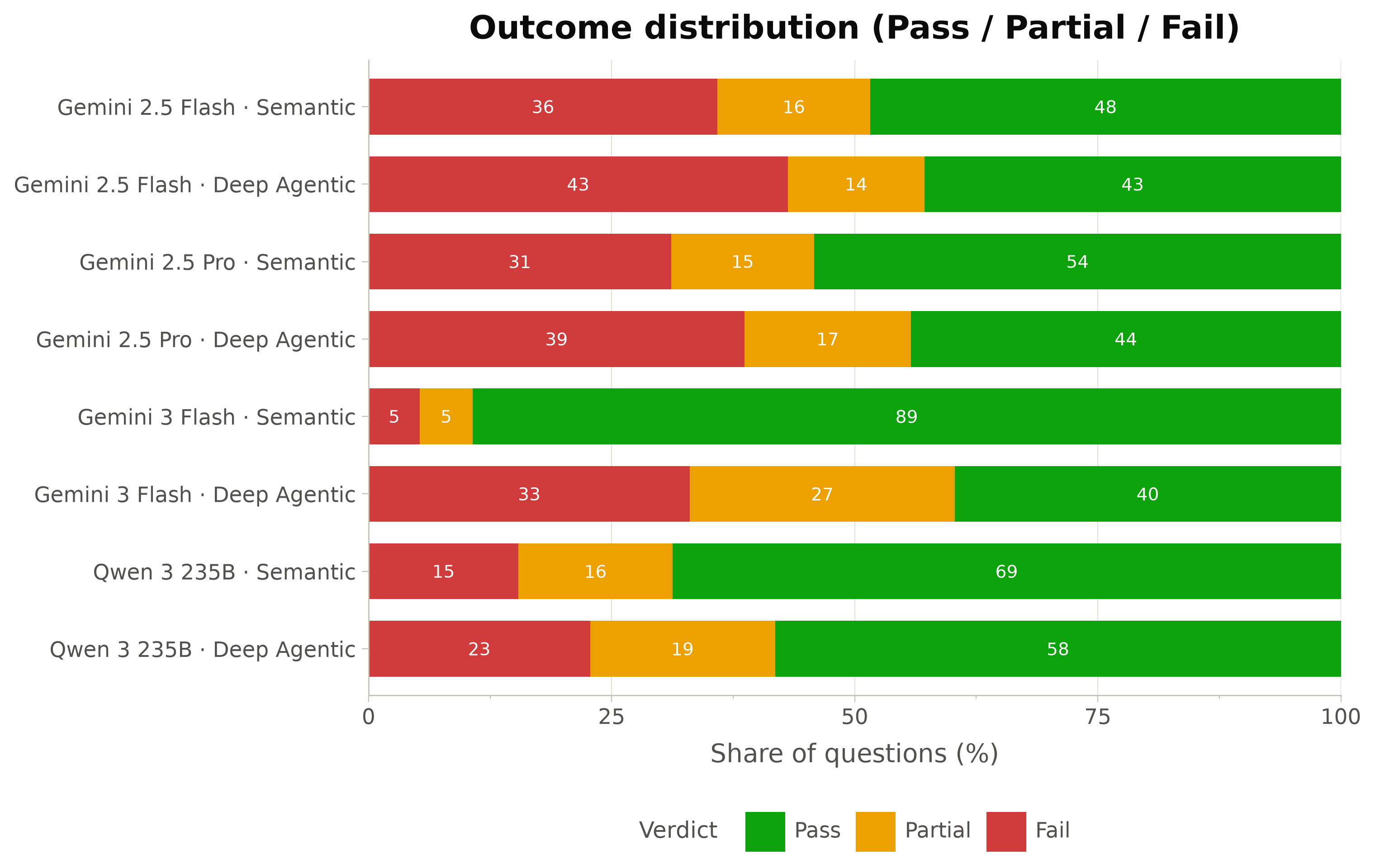}
   \caption{Verdict distribution for all eight conditions. Deep shifts mass from Pass toward Fail relative to Semantic for every model.}
   \label{fig:outcome}
\end{figure}

\begin{figure}[htbp]
   \centering
   \includegraphics[width=\textwidth]{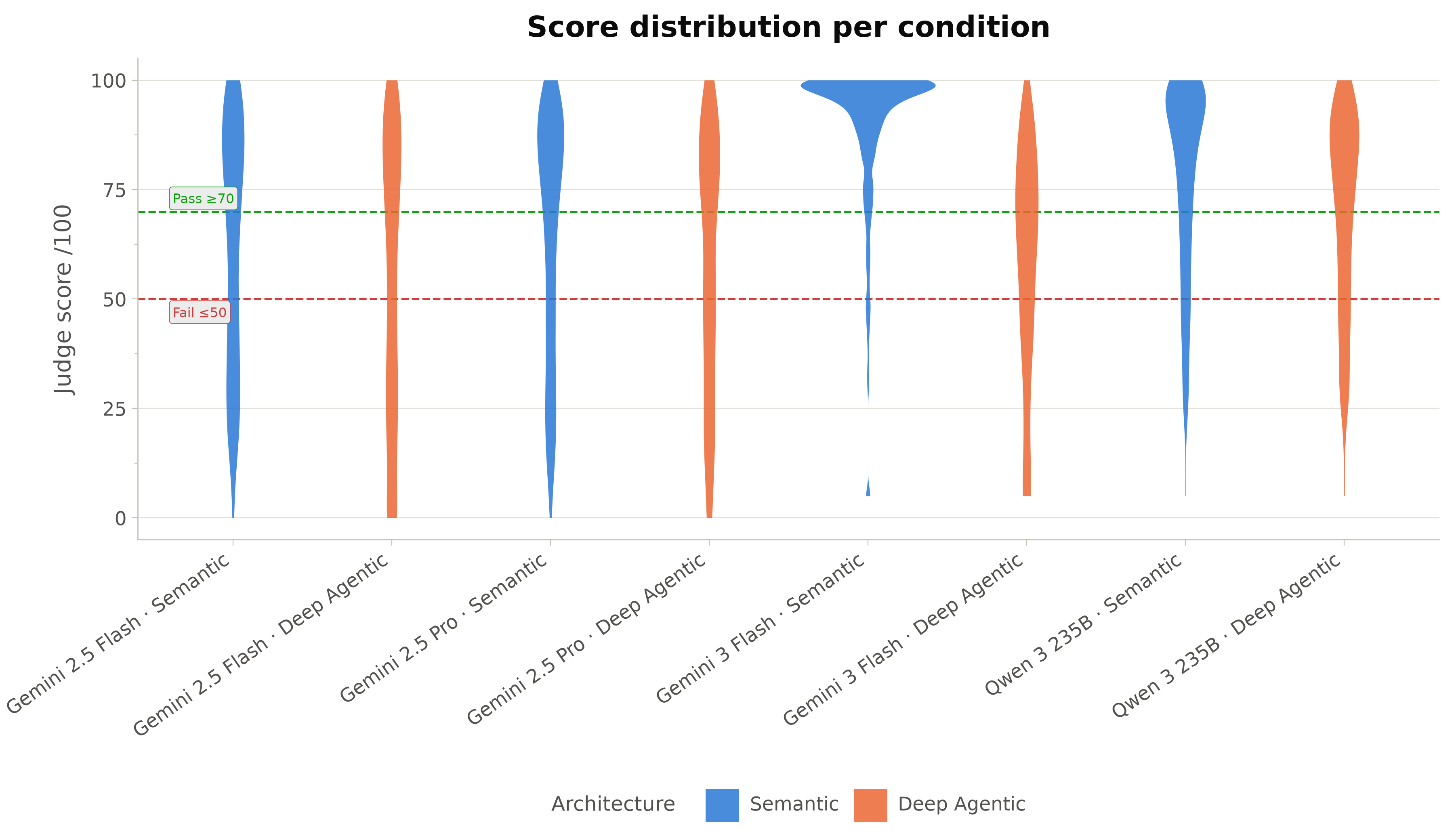}
   \caption{Judge score distribution per condition, with the Pass and Fail thresholds marked. Semantic concentrates mass at high scores, while Deep carries more mass in the low tail.}
   \label{fig:dist}
\end{figure}

The verdict distribution (Fig.~\ref{fig:outcome}) shows that Deep did not simply convert Pass answers into Partial answers; it also produced a larger share of outright failures for every model. The score distributions (Fig.~\ref{fig:dist}) make the same point at finer resolution: Semantic concentrates mass in the high score region, whereas Deep carries more mass in the low tail, most visibly for Gemini 3 Flash.

\begin{figure}[htbp]
   \centering
   \includegraphics[width=\textwidth]{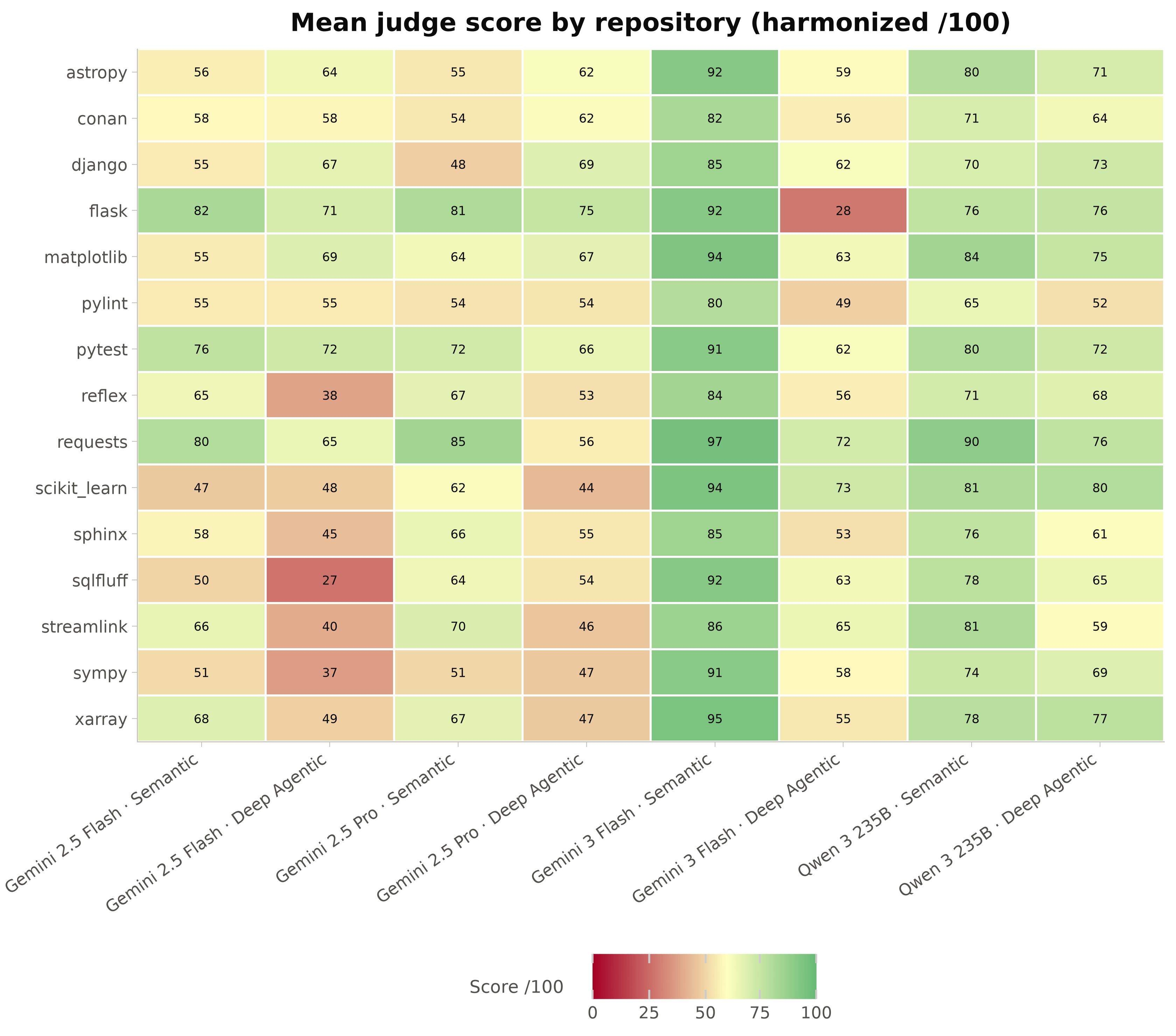}
   \caption{Mean judge score by repository for all eight conditions on a common 0 to 100 scale. Deep matched or exceeded Semantic on only one of the fifteen repositories.}
   \label{fig:repoheat}
\end{figure}

The advantage was not uniform across repositories (Fig.~\ref{fig:repoheat}). Pooling scores across the four models, Deep exceeded Semantic on only one of the fifteen repositories, django, by 3.0 points. Semantic led on the remaining fourteen, by margins of up to 22.9 points on streamlink. This heterogeneity indicates that Deep can be competitive on a particular repository, but that such cases were the exception rather than the rule in our sample. The advantage of Semantic did narrow as repository size grew, and across the fifteen repositories that trend was statistically detectable (Spearman's rho of $-0.69$, $p = 0.004$, against repository size in thousands of lines of Python). It did not reverse anywhere in our range: the largest repository in the sample, sympy at about 861 thousand lines of Python, still favoured Semantic by 13.7 points, and the single repository on which Deep led is mid-sized rather than large. We therefore read the size relationship as a narrowing margin rather than as a crossover, and note that establishing where, if anywhere, it crosses would need repositories larger than the ones studied here.

\begin{figure}[htbp]
   \centering
   \includegraphics[width=\textwidth]{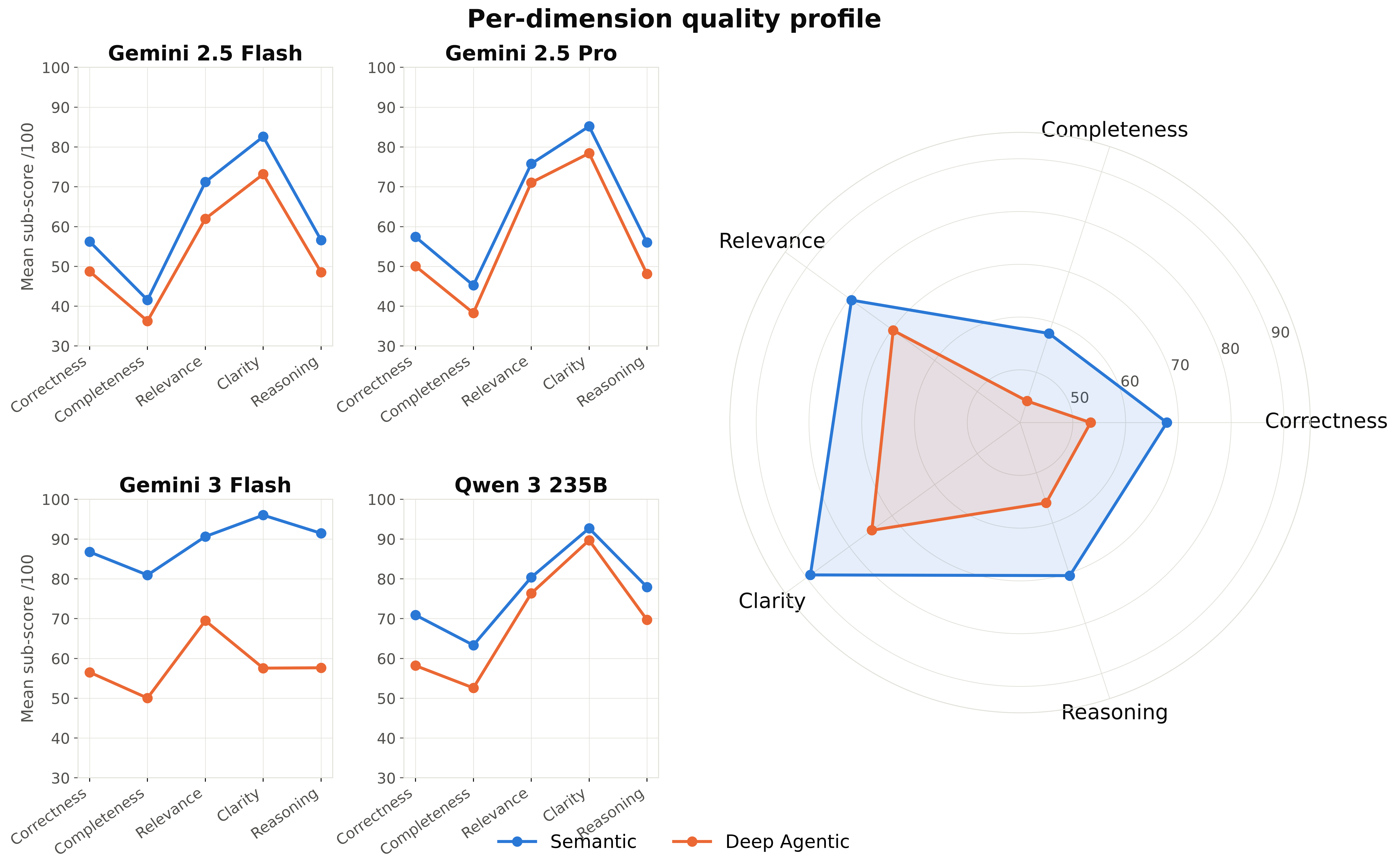}
   \caption{Per-dimension quality profile. Left: mean judge sub-score on each of the five dimensions for every model, comparing Semantic and Deep Agentic search. Right: the five dimensions pooled across the four models. Semantic is higher than Deep Agentic on every dimension.}
   \label{fig:dimprofile}
\end{figure}

Across the five quality dimensions (Fig.~\ref{fig:dimprofile}), Semantic scored higher than Deep on all five when pooled over the four models. The smallest gap was on relevance (about 10 points on the 0 to 100 scale) and the remaining four gaps were comparable, between about 13 and 15 points. Because the five dimensions carry equal weight in the total, we also examined correctness on its own, where Semantic led for every model, by between 7.3 and 30.2 points. Individual models showed some variation in which dimension separated the paradigms most, but no dimension favoured Deep on average.

\begin{figure}[htbp]
   \centering
   \includegraphics[width=\textwidth]{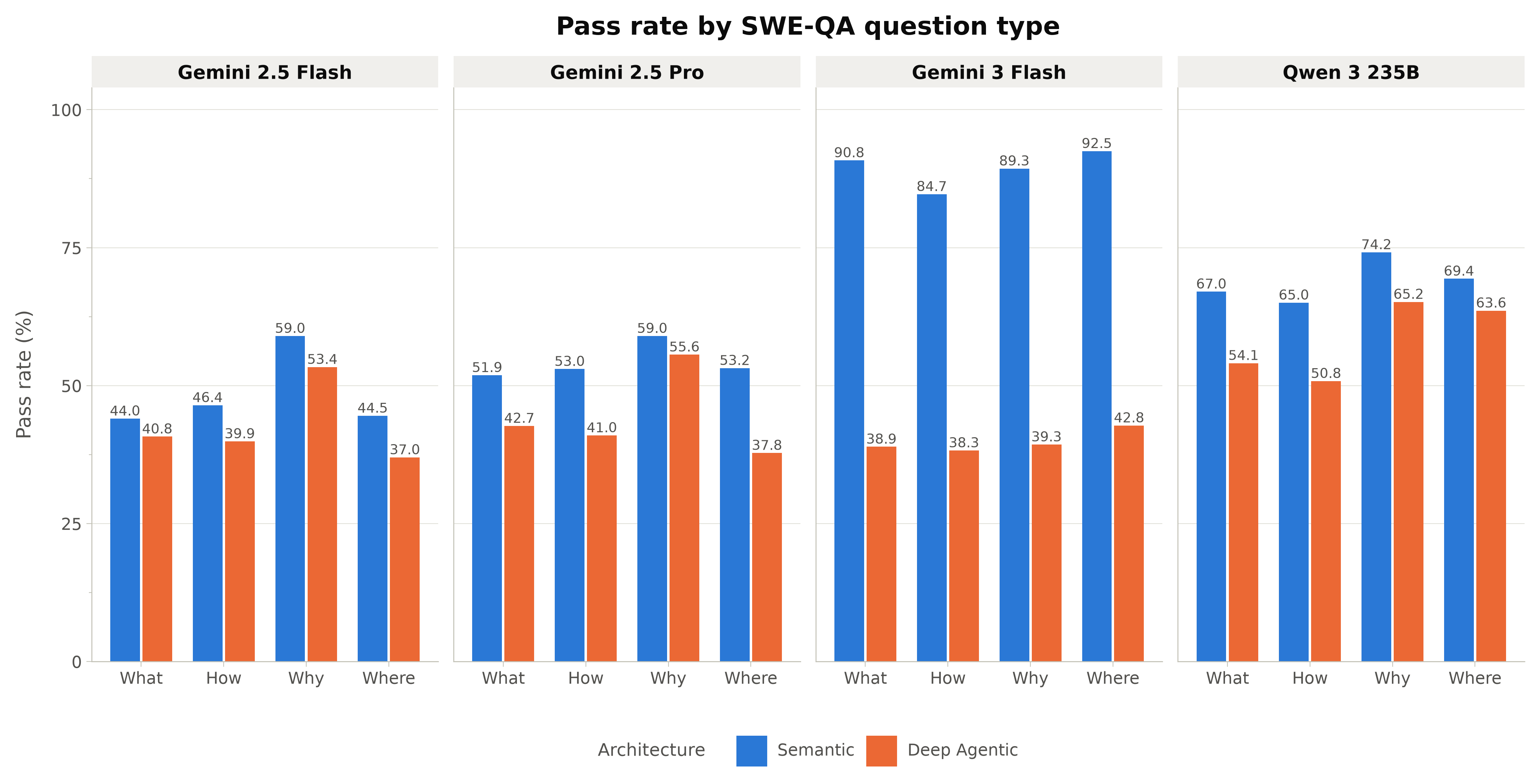}
   \caption{Pass rate by question type for each model. Semantic exceeds Deep for every question type, with a roughly uniform gap.}
   \label{fig:qtype}
\end{figure}

Finally, the advantage held across all four question types (Fig.~\ref{fig:qtype}). Pooled over models, the Semantic Pass rate exceeded the Deep Pass rate for What, Why, Where, and How questions, with a fairly uniform gap of roughly 17 to 20 percentage points. Both paradigms found Why questions easiest and How questions hardest, which suggests that question type shifts the absolute difficulty for both paradigms without changing which paradigm is preferable.

\subsection{Computational cost and efficiency}
\label{sec:rq2}
Our second research question concerns computational and monetary cost. Deep was more expensive than Semantic on the token and dollar measures for every model (Table~\ref{tab:master}). Deep consumed more input tokens than Semantic in all four cases; the difference was smallest for the Gemini 2.5 pair and largest for Qwen3-235B, where Semantic used about 34 thousand input tokens per question against about 761 thousand for Deep. The mean dollar cost per question was higher for Deep than for Semantic for every model.

\begin{figure}[htbp]
   \centering
   \includegraphics[width=\textwidth]{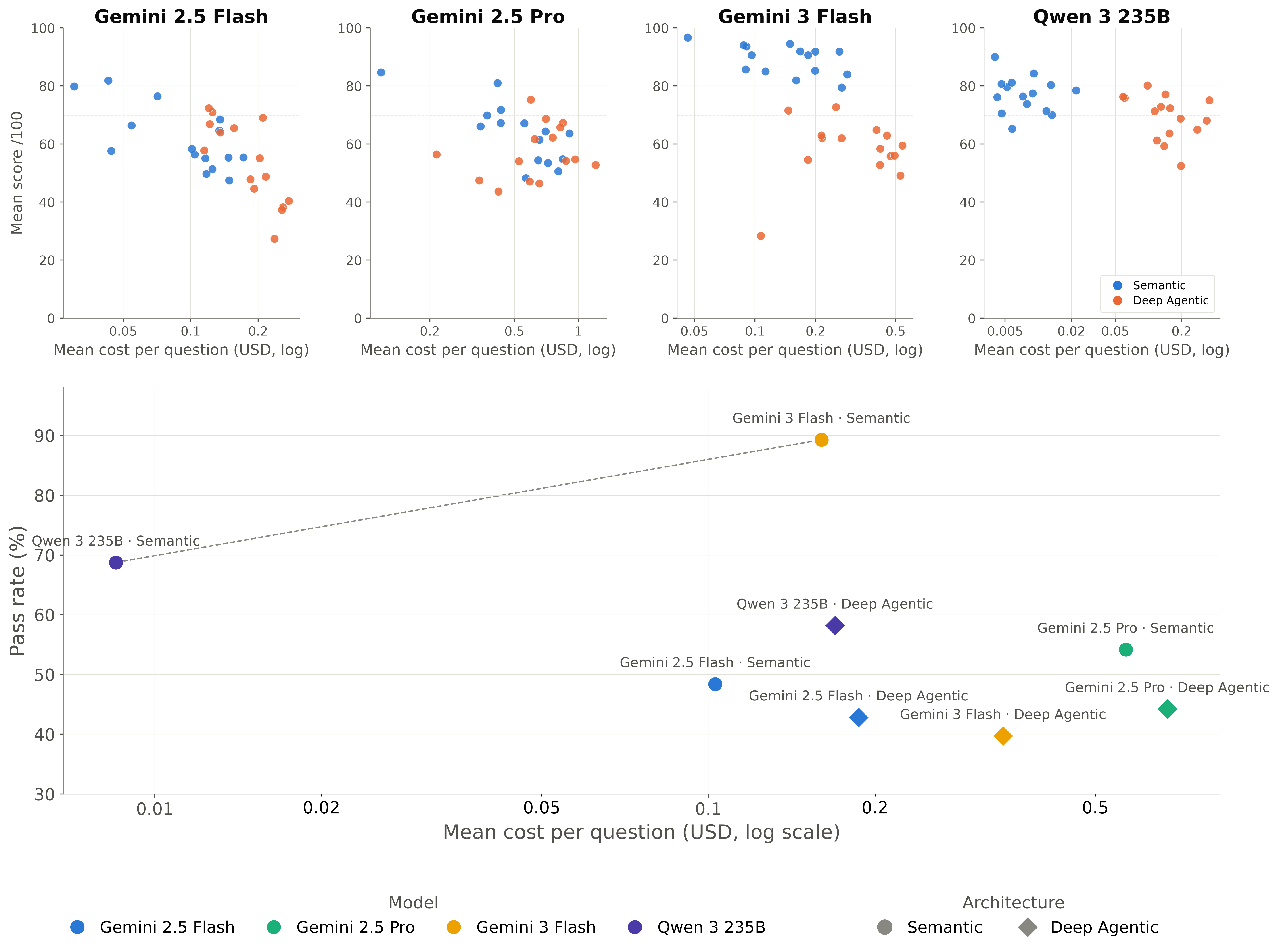}
   \caption{Accuracy against cost. Bottom: the eight conditions in the accuracy and cost plane; Semantic conditions occupy the higher-accuracy, lower-cost region and the dashed line marks the efficient frontier. Top: the same trade-off at the repository level, one panel per model, each point a repository (dashed line marks Pass at 70).}
   \label{fig:pareto}
\end{figure}

\begin{figure}[htbp]
   \centering
   \includegraphics[width=0.82\textwidth]{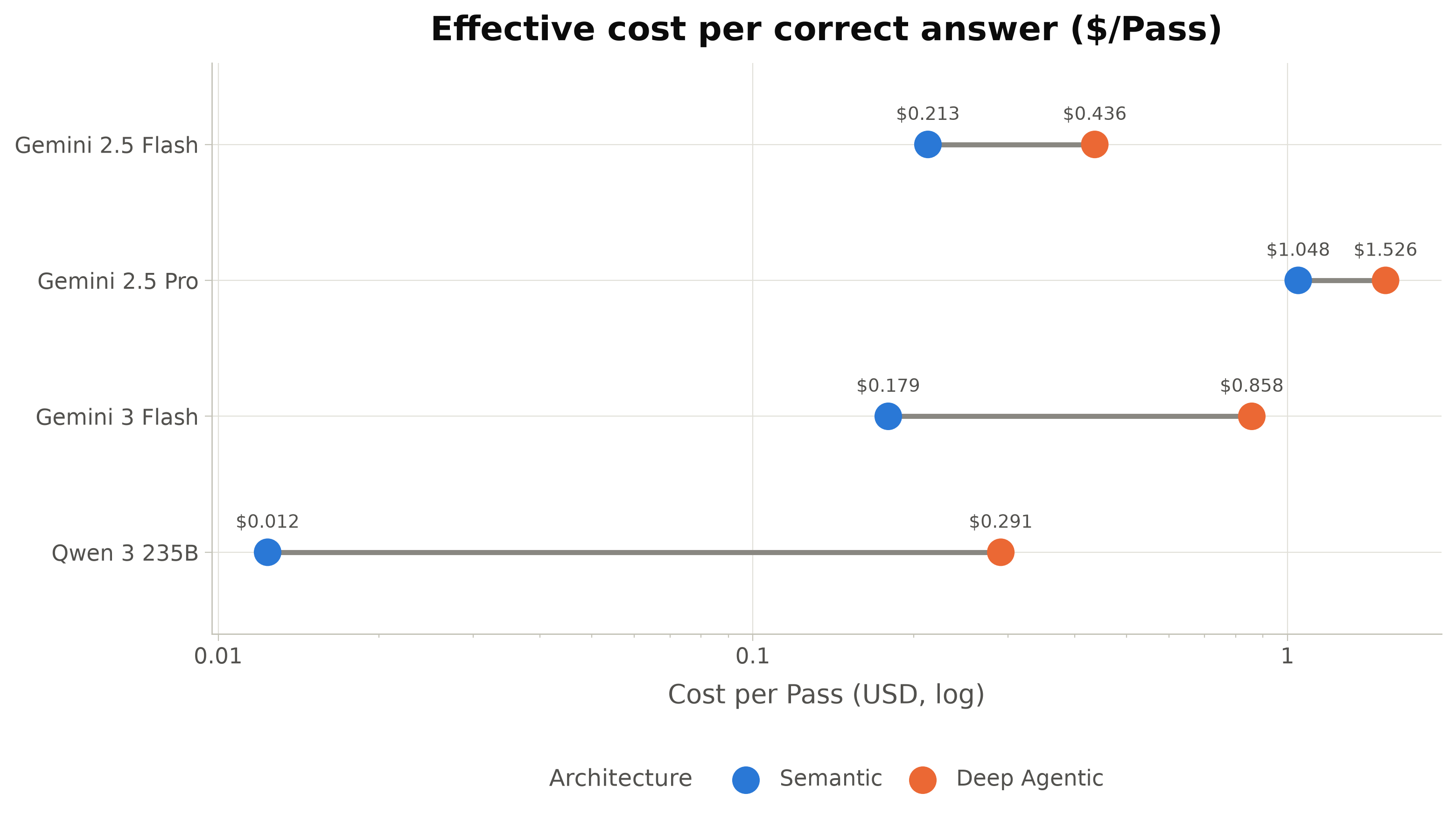}
   \caption{Effective cost per correct answer. For every model, the Deep point sits at a higher cost per Pass than the Semantic point.}
   \label{fig:costpass}
\end{figure}

Because Deep was both less accurate and more costly, it did not lie on the efficient frontier of the accuracy against cost trade-off. Figure~\ref{fig:pareto} places the eight conditions in the accuracy and cost plane: the Semantic conditions occupy the upper-left region of higher accuracy at lower cost, while the Deep conditions sit lower and to the right. The same pattern appeared at the repository level (Fig.~\ref{fig:pareto}), where the Semantic points for each model clustered at lower cost than the corresponding Deep points. The practitioner-facing view, the cost per correct answer, tells the same story (Fig.~\ref{fig:costpass}): for every model, Deep cost more per Pass than Semantic, for example 0.179 against 0.858 dollars for Gemini 3 Flash and 0.012 against 0.291 dollars for Qwen3-235B. Pooled across the four models, a correct answer cost 0.32 dollars with Semantic against 0.74 dollars with Deep, so Deep was about 2.3 times more expensive for each question it answered correctly.

Latency did not order the two paradigms consistently. The median latency was lower for Deep than for Semantic for the Gemini 2.5 pair and higher for Deep for Gemini 3 Flash and Qwen3-235B (Table~\ref{tab:master}). Latency was also heavy-tailed and was affected by the execution environment, so we treat tokens and dollar cost as the primary cost measures and interpret latency cautiously.

\begin{figure}[htbp]
   \centering
   \includegraphics[width=\textwidth]{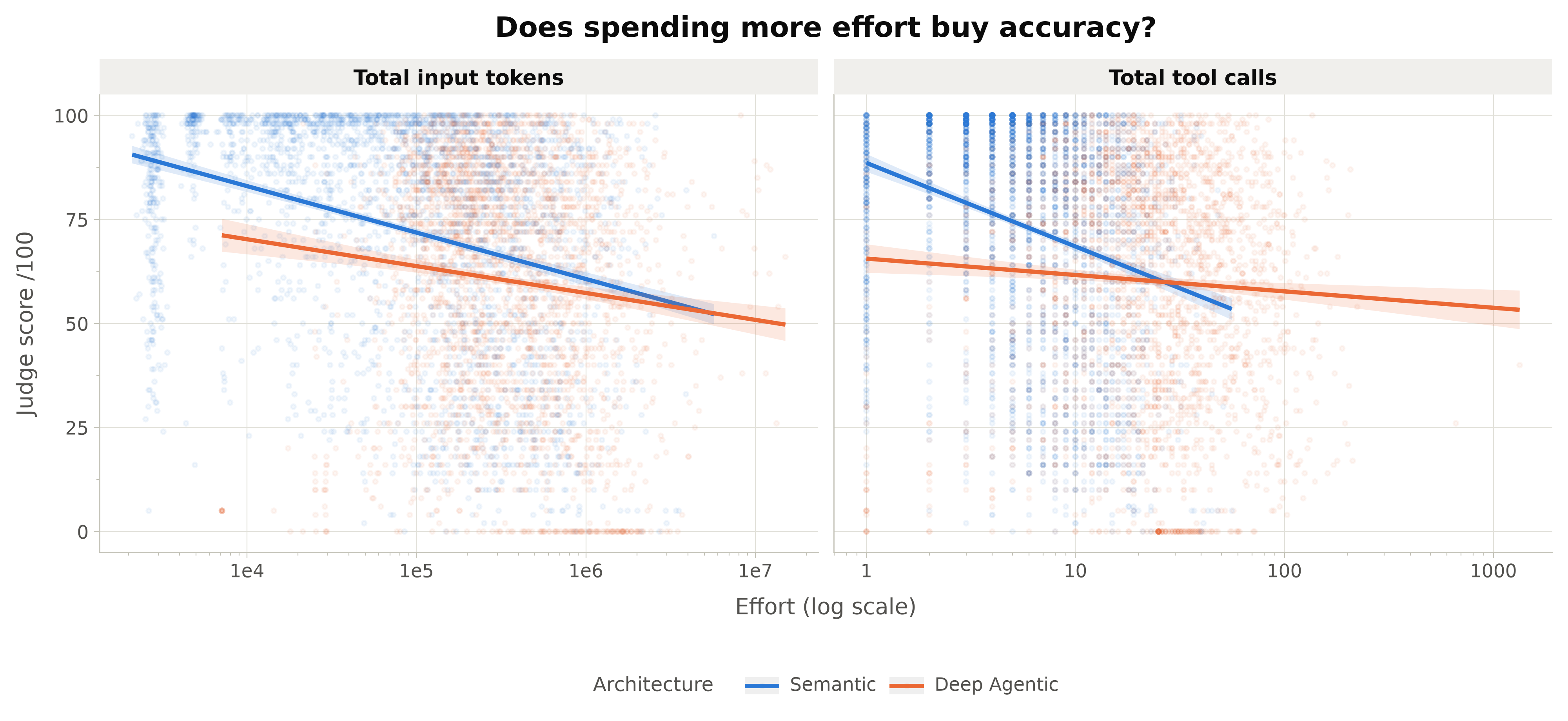}
   \caption{Per-question score against effort, measured by input tokens and by tool calls, with ordinary least squares fits. Slopes are flat to negative for both paradigms.}
   \label{fig:effort}
\end{figure}

A natural question is whether the extra effort that Deep spends is simply the price of harder questions, or whether it buys accuracy. Figure~\ref{fig:effort} plots the per-question score against the amount of effort, measured by total input tokens and by total tool calls. For both paradigms the fitted slope is flat to negative: spending more tokens or making more tool calls was not associated with higher scores. This relationship is correlational, and it partly reflects that harder questions require more effort and also score lower, so we do not interpret it causally. It does show, however, that within our data additional effort was not accompanied by additional accuracy.

\begin{figure}[htbp]
   \centering
   \includegraphics[width=\textwidth]{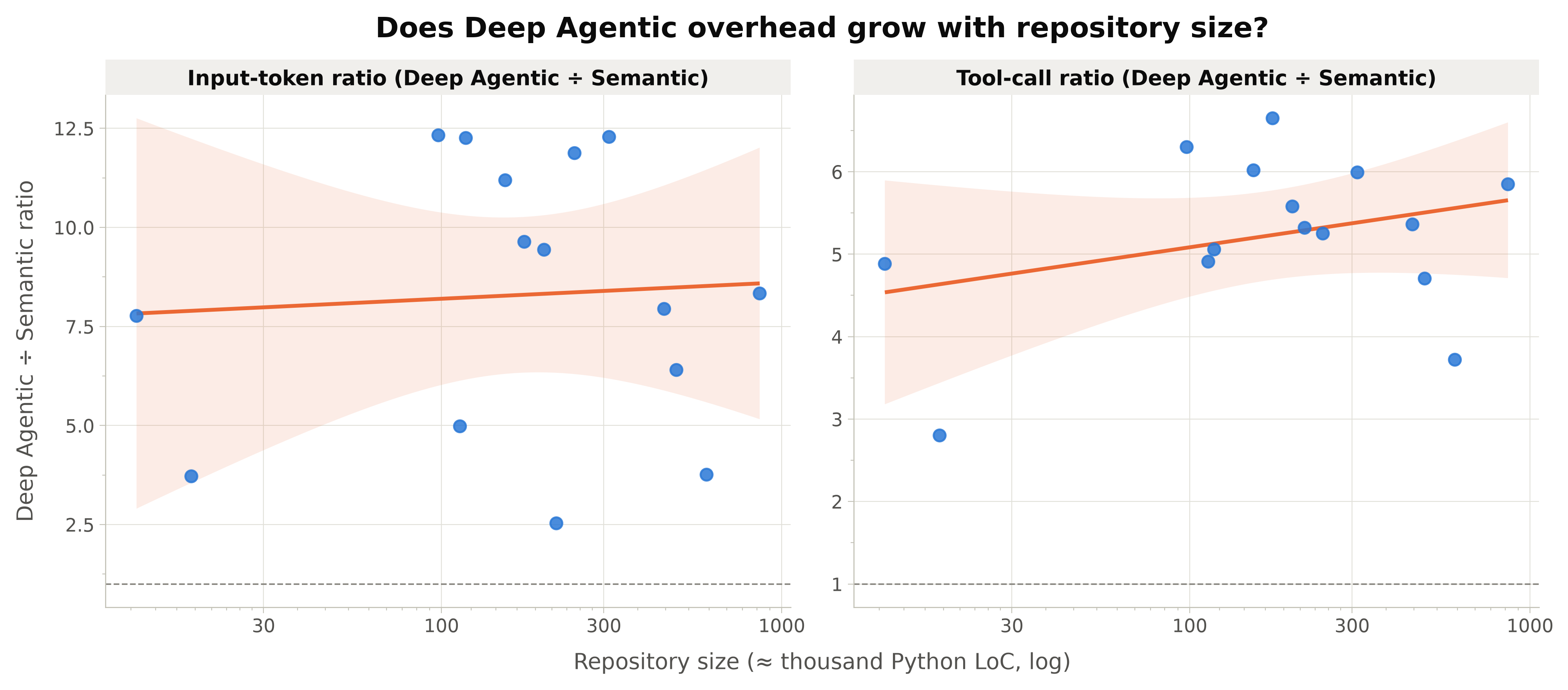}
   \caption{Ratio of Deep to Semantic per-question effort against repository size, pooled over models. No significant association is detected, though with fifteen repositories the test is underpowered and the result is inconclusive.}
   \label{fig:scaling}
\end{figure}

We also tested whether the cost gap grows with repository size, since one might expect dynamic exploration to scale worse than a pre-built index on larger repositories. Figure~\ref{fig:scaling} plots the per-question ratio of Deep to Semantic effort against repository size. We found no significant association between the overhead ratio and repository size, but with only fifteen repositories the interval is wide, so we treat this as inconclusive rather than as evidence of no relationship. What the data do show is that the Deep overhead was present at every repository size in our sample rather than appearing only on the largest repositories.

\subsection{Agent behaviour}
\label{sec:rq3}
Our third research question concerns how the two architectures behave and how they fail. This subsection reports the two behavioural measures, tool usage and trajectory length; the analysis of how and why the agents fail follows in Section~\ref{sec:failure}.

\begin{figure}[htbp]
   \centering
   \includegraphics[width=\textwidth]{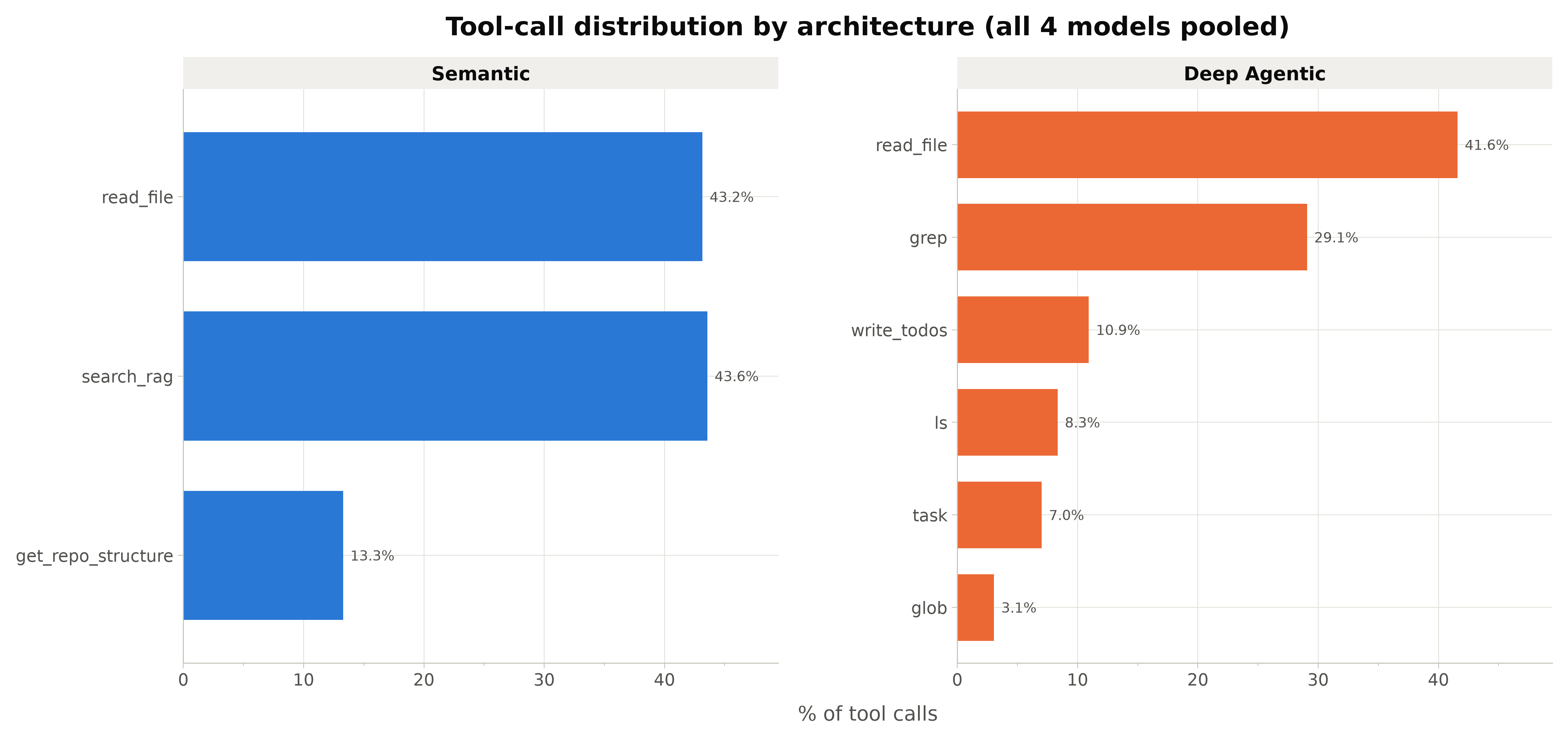}
   \caption{Tool-call distribution by paradigm, pooled across the four models, counting the calls of the delegated sub-agent alongside the planner's own. Deep spreads its calls across more tools but is dominated by reading and searching. Delegation is 7.0\% of its calls, yet those calls open the sub-agent trajectories in which 68.8\% of all Deep tool calls are made.}
   \label{fig:tooluse}
\end{figure}

\paragraph{Tool usage.}
The tool-usage distribution (Fig.~\ref{fig:tooluse}) shows how each architecture spent its actions. Semantic distributed its calls across only three tools, principally retrieval (43.6\%) and file reading (43.2\%), with the remainder on structure lookup (13.3\%). Deep had a larger toolset, and once the actions of its delegated sub-agent are counted alongside the planner's own, the great majority of its activity was ordinary reading and searching: file reading (41.6\%) and text search (29.1\%) together accounted for more than two thirds of its calls, followed by task-list writing (10.9\%), directory listing (8.3\%), sub-agent delegation (7.0\%), and file globbing (3.1\%).

Counting delegations as a share of all calls understates how much of the work they govern, because a delegation is a single call that opens an entire sub-agent trajectory. Of the 2,878 Deep runs, 2,395 delegated at least once, 2.1 times per run on average, and each delegation ran about ten further calls of its own. Taken together, 68.8\% of all Deep tool calls were issued inside a delegated sub-agent rather than by the planner. Measured within the planner's own context, delegation was its second most frequent action at 22.4\%, behind only task-list writing. The deep agent therefore spent most of its budget on the same reading and searching operations that are available in principle to a simpler agent, but it spent it at one remove, with more than two thirds of that activity taking place in a context the planner never saw directly. Section~\ref{sec:failure} returns to this point, because the hand-off that governs those two thirds is also where the largest share of Deep failures occurred.

\begin{figure}[htbp]
   \centering
   \includegraphics[width=0.9\textwidth]{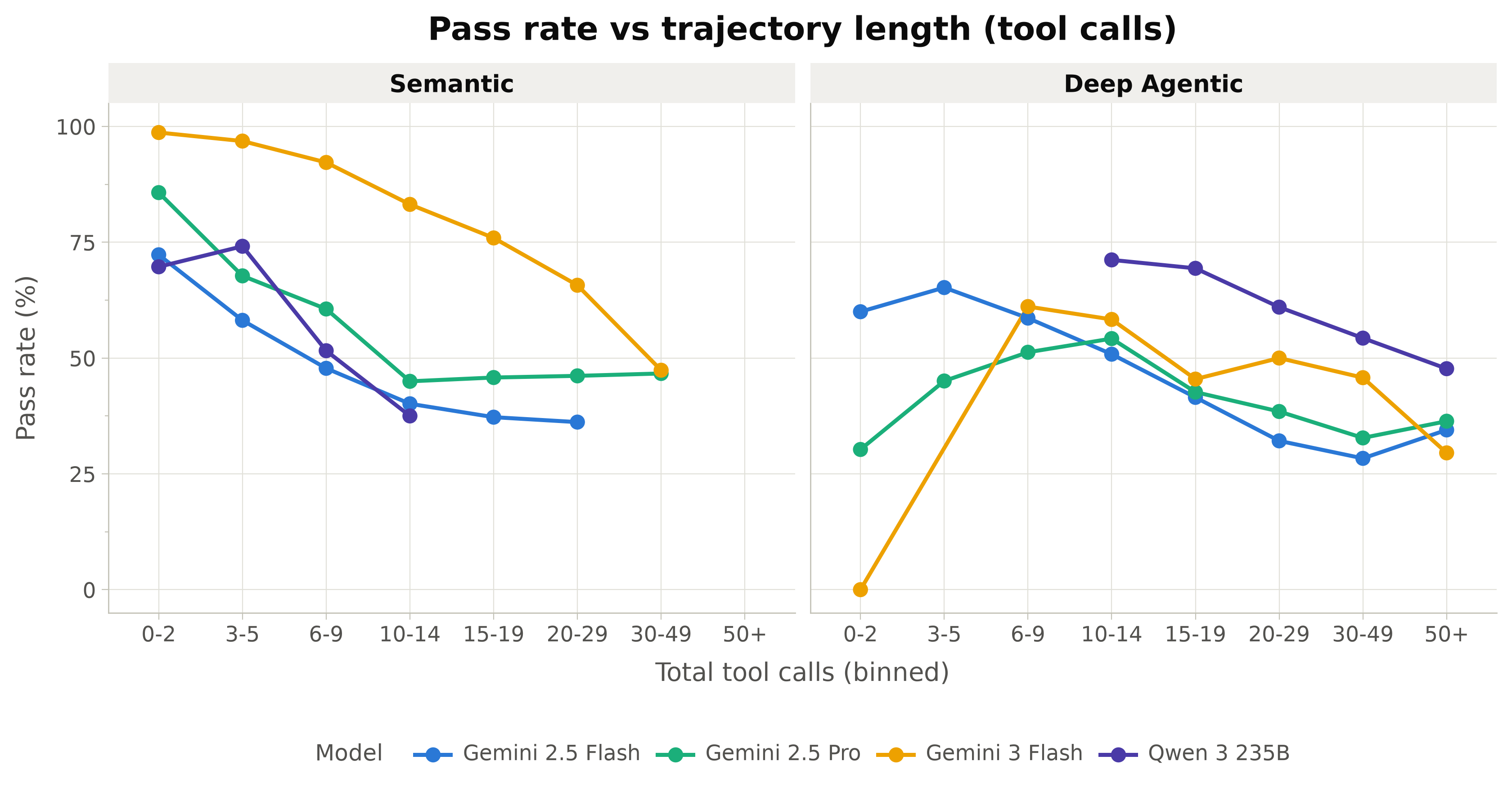}
   \caption{Pass rate against the number of tool calls per question. Bins containing fewer than ten questions are omitted as too sparse to interpret. In both paradigms, longer trajectories were associated with lower Pass rates.}
   \label{fig:traj}
\end{figure}

\paragraph{Trajectory length and success.}
The relationship between trajectory length and success (Fig.~\ref{fig:traj}) followed a diminishing-returns pattern in both architectures. Once a run was under way, the Pass rate fell as the number of tool calls per question grew, and it did so over a much longer range for Deep, whose trajectories extend well past fifty calls once the actions of its sub-agent are counted. The very shortest Deep trajectories are the exception, and they are low for a different reason: they are dominated by runs that stalled before doing any real work. Away from that end, longer trajectories were associated with lower rather than higher success, which is consistent with the reading that questions requiring extended exploration are also the harder questions and that extended exploration did not recover them.

\section{Failure Analysis}
\label{sec:failure}
This section answers the remainder of our third research question, namely how and why each architecture fails. We coded every failure and then examine the resulting distributions and what they reveal.

\subsection{Failure Analysis Methodology}
We coded every failure, defined as a run with a Fail verdict, on two orthogonal axes. The mechanism axis records the root behavioural cause of the failure, and the symptom axis records what was wrong with the answer, so that the mechanism does not depend on the answer content and the symptom does not depend on how the answer was produced. Each failure was distilled from its recorded trace into a compact record holding the question, the reference answer, the judge's rationale, the agent's answer, and the cleaned tool sequence, and every record was then coded on both axes by Claude Sonnet 4.6. To validate this coding, two of the authors independently coded a stratified random sample of 400 failures, about a quarter of the failure set. Agreement between the two authors, measured by Cohen's kappa, was 0.86 on the symptom (90.5\% raw agreement) and 0.78 on the mechanism (83.5\% raw agreement), and agreement between their consensus labels and the automatic coding was 0.82. On the scale of \citet{landis1977} these values are substantial to almost perfect. Disagreements between the authors were resolved by discussion to consensus after review of the full execution traces.

\subsection{Failure taxonomy}
The taxonomy has six mechanism codes and five symptom codes. The mechanisms are a coordination breakdown, in which the planner to sub-agent hand-off fails to produce or integrate a usable result and which is possible only for the Deep Agent; a retrieval or localization miss, in which substantial search and read effort never surfaces the correct file or entity; insufficient grounding, in which a confident answer is produced after little or no reading; evidence misinterpreted, in which the agent reads the correct file but misreports or overreaches in synthesis; a non-terminating loop, in which the control loop cycles or stalls without producing an answer; and an evidence-gathering failure, in which tool calls return errors and the run never obtains the material it needs. The symptoms are no answer; premise denial, a claim that the requested code does not exist; a wrong entity or mechanism, a confident answer about the wrong file or mechanism; factually incorrect, a wrong statement about the right entity; and incomplete or shallow, an on-topic answer that omits most of the required detail.

\subsection{Failure distributions}
\begin{figure}[htbp]
   \centering
   \includegraphics[width=0.82\textwidth]{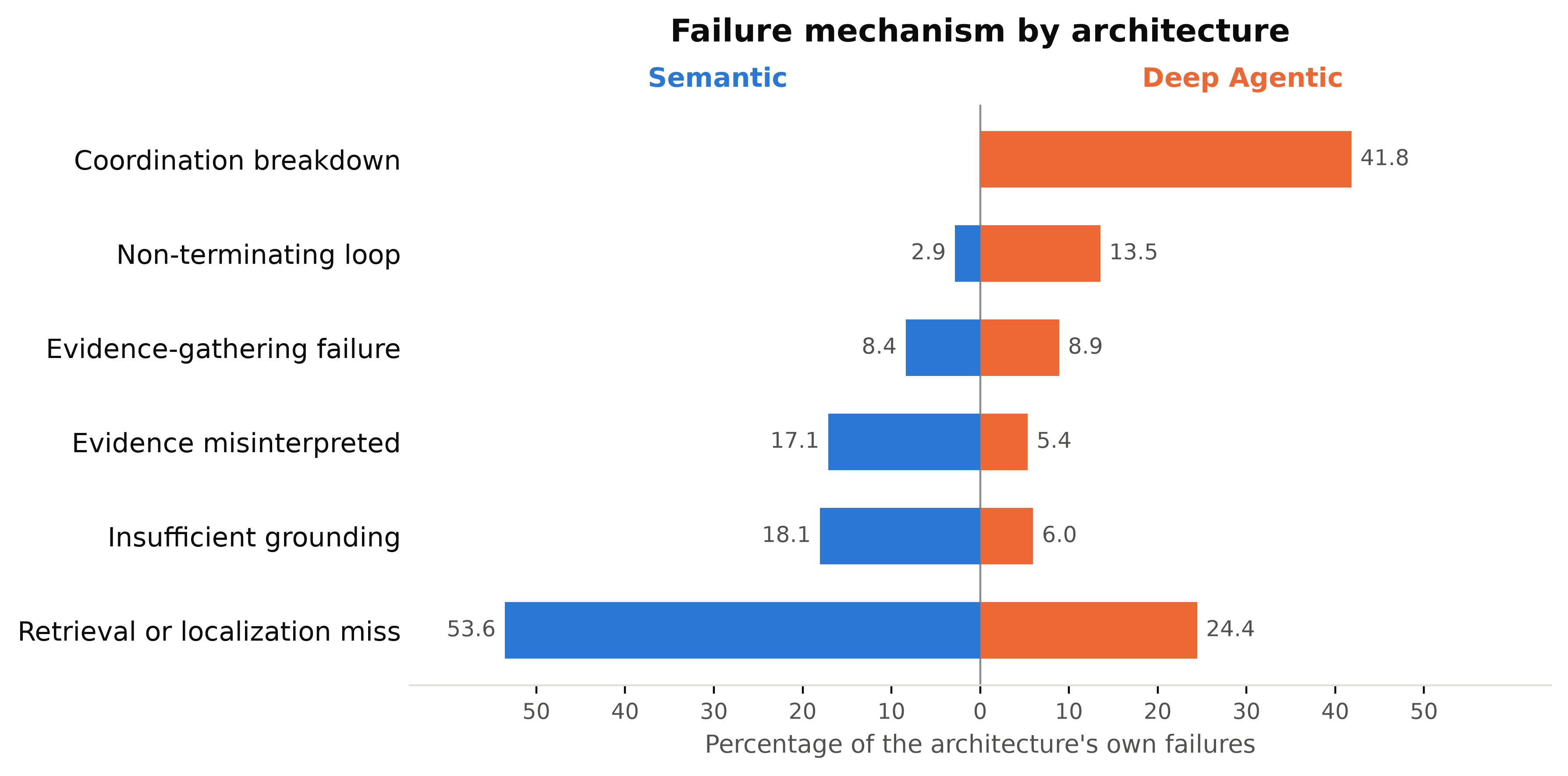}
   \caption{Failure mechanism by architecture, as a percentage of each architecture's own failures. Semantic is on the left and Deep on the right. The two architectures fail through different mechanisms.}
   \label{fig:mecharch}
\end{figure}

\begin{figure}[htbp]
   \centering
   \includegraphics[width=0.82\textwidth]{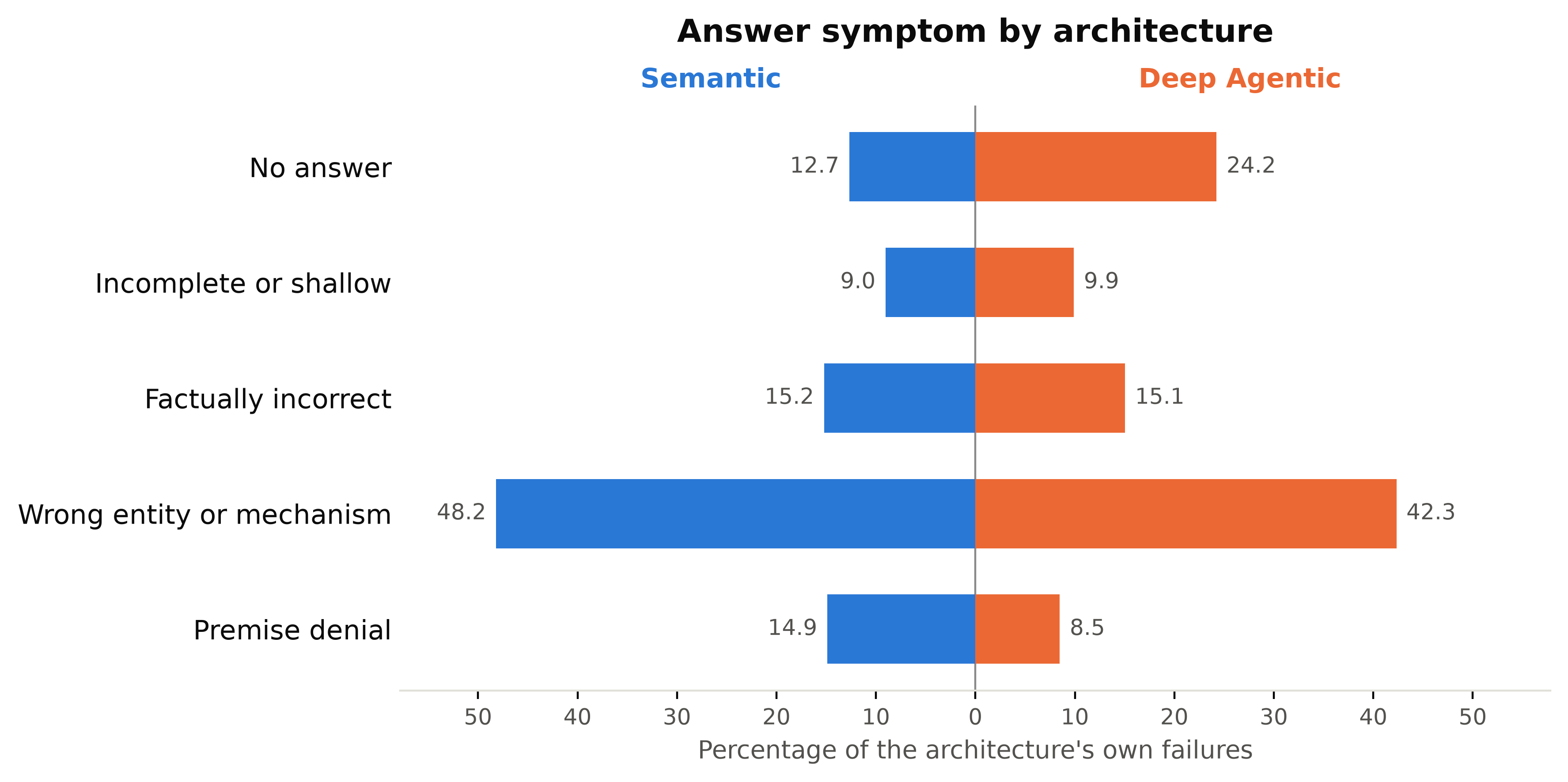}
   \caption{Answer symptom by architecture, as a percentage of each architecture's own failures. The symptom differs between the architectures far less than the mechanism does.}
   \label{fig:symarch}
\end{figure}

\begin{figure}[htbp]
   \centering
   \includegraphics[width=\textwidth]{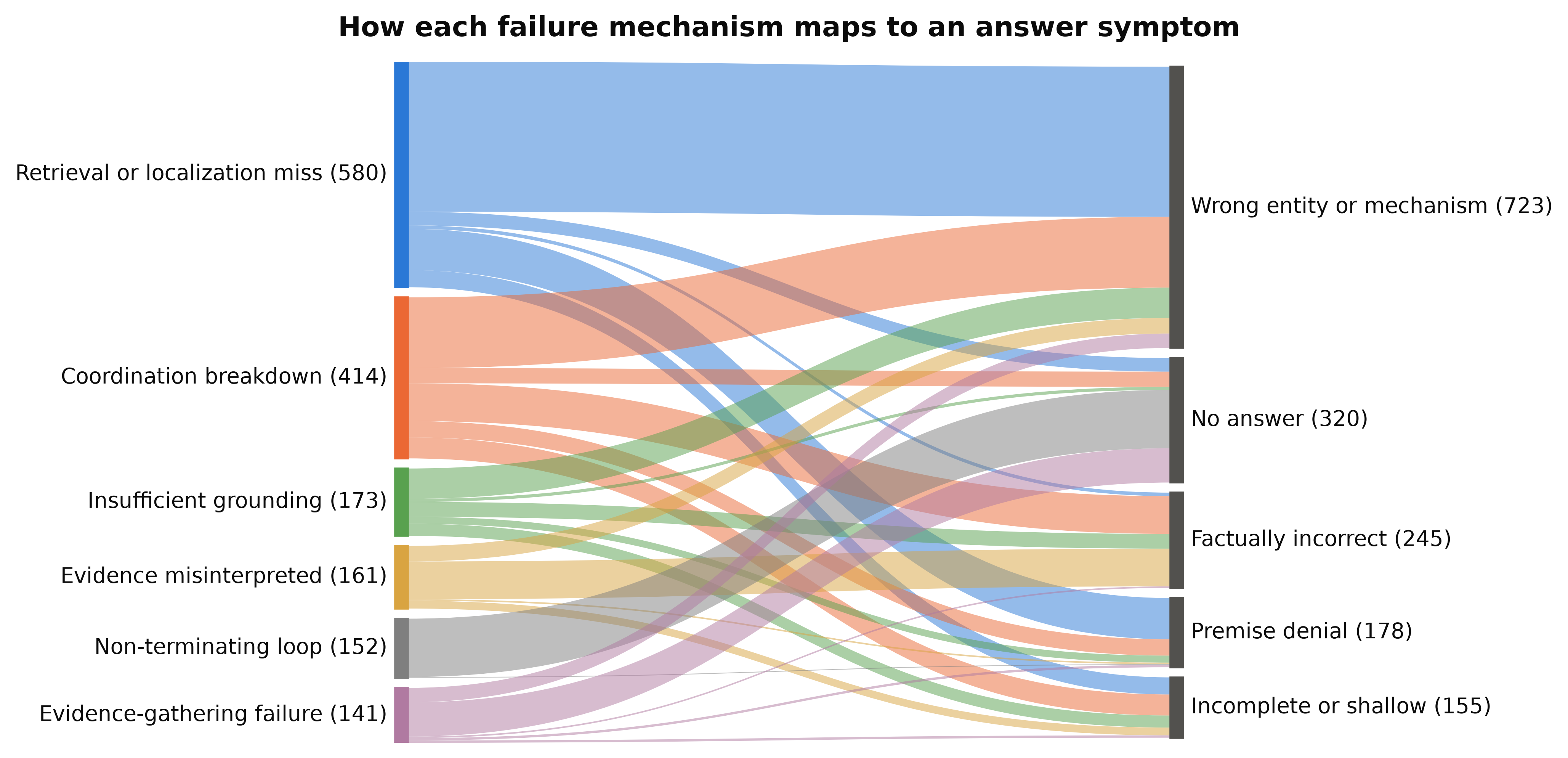}
   \caption{How each failure mechanism maps to an answer symptom, pooled across both architectures. Ribbon width is proportional to the number of failures.}
   \label{fig:sankey}
\end{figure}

Figure~\ref{fig:mecharch} reports the failure mechanism by architecture. The association is very strong, with a chi-square of 508.8 on five degrees of freedom and a Cramer's $V$ of 0.56. Part of that association is definitional, because a coordination breakdown can only arise in a hierarchical design. Restricting the comparison to the five mechanisms available to both architectures removes that component, and the association remains large, with a chi-square of 147.2 on four degrees of freedom and a Cramer's $V$ of 0.35. The largest share of Deep failures is the coordination breakdown, at 41.8\% of them and none for Semantic, because a flat retrieval agent has no planner to sub-agent hand-off that can break. Semantic failures instead concentrate in retrieval and localization. The symptom is associated with the architecture much less strongly (Fig.~\ref{fig:symarch}, chi-square 43.8, Cramer's $V$ of 0.16), so the two architectures produced wrong answers of broadly similar shape while reaching them through different mechanisms. Figure~\ref{fig:sankey} traces which mechanism leads to which symptom, and the mapping is close to deterministic for several mechanisms: the non-terminating loop flows almost entirely to no answer, while the coordination breakdown fans mostly to a confident but wrong entity. Table~\ref{tab:failall} gives the full numbers, including the severity of each mechanism, measured by the mean judge score among its failures. Two mechanisms often end without a usable answer. The non-terminating loop scores close to zero, at a mean of 3.3, and the evidence-gathering failure scores higher at 18.8 but still well below the content mechanisms, which cluster near thirty.

\begin{table}[htbp]
   \centering
   \small
   \caption{Full failure-analysis results. Upper block: the six failure mechanisms, each as a percentage of the architecture's own failures with counts in parentheses, together with the pooled share of that mechanism that produced no answer and its mean judge score. Lower block: the five answer symptoms by architecture. The failure set contains 1{,}621 failures, 631 for Semantic and 990 for Deep. Mechanism by architecture: $\chi^2 = 508.8$, Cramer's $V = 0.56$; over the five mechanisms available to both architectures, $\chi^2 = 147.2$, Cramer's $V = 0.35$. Symptom by architecture: $\chi^2 = 43.8$, Cramer's $V = 0.16$. Agreement on a stratified 400-failure subsample: Cohen's kappa 0.86 (symptom) and 0.78 (mechanism) between the two author annotators, and 0.82 between their consensus and the automatic coding.}
   \label{tab:failall}
   \begin{tabular}{lrrrr}
      \toprule
      Mechanism                      & Semantic   & Deep       & No answer (\%) & Mean score \\
      \midrule
      Coordination breakdown         & 0.0 (0)    & 41.8 (414) & 9.4            & 30.5       \\
      Retrieval or localization miss & 53.6 (338) & 24.4 (242) & 6.0            & 29.0       \\
      Insufficient grounding         & 18.1 (114) & 6.0 (59)   & 4.6            & 33.3       \\
      Evidence misinterpreted        & 17.1 (108) & 5.4 (53)   & 0.0            & 38.4       \\
      Non-terminating loop           & 2.9 (18)   & 13.5 (134) & 98.7           & 3.3        \\
      Evidence-gathering failure     & 8.4 (53)   & 8.9 (88)   & 62.4           & 18.8       \\
      \bottomrule
   \end{tabular}

   \vspace{0.7em}

   \begin{tabular}{lrr}
      \toprule
      Symptom                   & Semantic   & Deep       \\
      \midrule
      Wrong entity or mechanism & 48.2 (304) & 42.3 (419) \\
      No answer                 & 12.7 (80)  & 24.2 (240) \\
      Factually incorrect       & 15.2 (96)  & 15.1 (149) \\
      Premise denial            & 14.9 (94)  & 8.5 (84)   \\
      Incomplete or shallow     & 9.0 (57)   & 9.9 (98)   \\
      \bottomrule
   \end{tabular}
\end{table}

\subsection{Key findings}
\paragraph{The Semantic search signature.}
Semantic failures were about finding and reading the right code rather than reasoning about it. Its largest mechanism was a retrieval or localization miss, in 53.6\% of Semantic failures, and these were high-effort misses, with a mean of 12.6 tool calls spent on searches that never resolved to the right file. A further 18.1\% answered on too little evidence, and 17.1\% read the genuinely correct file but misread it, the last of these reading roughly four times as many files as a retrieval miss, which indicates that the correct material was in the context and was misread rather than never found. At the level of the answer, Semantic was also more likely than Deep to declare that the requested code did not exist, in 14.9\% of its failures against 8.5\%, though this gap was not uniform across the four models.

\paragraph{The Deep Agent signature.}
Deep failures were dominated by the coordination machinery itself. Coordination breakdown accounted for 41.8\% of Deep failures and has no counterpart in a flat retrieval agent, and a further 13.5\% were non-terminating loops. Deep was also far more likely to return nothing at all, in 24.2\% of its failures against 12.7\% for Semantic. The coordination breakdown itself, however, was usually not a visible blank: in 91\% of these cases the agent still returned a fluent, confidently worded answer rather than reporting that anything had gone wrong, so the breakdown was not apparent from the output alone. A hierarchical delegation that yields fluent but ungrounded answers of this kind is consistent with the general difficulty of transferring decisions and context across an agent-to-agent hand-off, rather than with a reasoning error about the code.

\paragraph{Silent and visible failures.}
The two signatures also differ in how visible they are. Semantic tended to fail with a fluent, confident, and wrong answer, whereas Deep failed either with such a confident wrong answer from a coordination breakdown or, more often than Semantic, with no answer at all. The deep architecture therefore did not remove failures; it added a class of failure, in its coordination, that a flat retrieval agent cannot have, and that failure is often silent.

\section{Discussion}
\label{sec:discussion}
Our central result runs against the prevailing direction of practice. For read-only repository-level code question answering, across four models and fifteen repositories, an index-backed Semantic search agent matched or exceeded a delegation-based Deep Agentic search agent in accuracy while costing less. The momentum in industry is toward more elaborate Deep Agentic search, so a finding that the simpler and older paradigm is at least as accurate calls for an explanation rather than a restatement of the numbers.

We read the failure analysis as an account of why the additional machinery did not pay off. The premise of Deep Agentic search is a form of context engineering \citep{anthropic2025context}: a sub-agent absorbs the noisy exploration and returns a condensed result, so that the orchestrator can reason over clean material. That same hand-off, however, is where the paradigm most often broke in our data. The dominant Deep failure was a coordination breakdown that has no analogue in a flat agent, and it was usually silent, producing a confident but wrong answer rather than a visible error. This is consistent with a growing argument among practitioners that multi-agent systems are fragile \citep{cognition2025multiagents} precisely because decisions and context do not transfer reliably across an agent-to-agent boundary. On a task whose difficulty is dominated by locating and correctly reading the right code, introducing that boundary added a class of failure that outweighed whatever it was meant to prevent.

Accuracy, however, is not the only reason a practitioner would reach for Deep Agentic search, and our study is not positioned to observe the others. The purpose of the paradigm is to keep the main reasoning context uncluttered by delegating the raw exploration elsewhere. In a modern agentic workflow a session is long-running and spans many tasks, and a developer cannot reset the context after every question without discarding the state that has accumulated over the session. Under those conditions, keeping the orchestrator's context clean has value in its own right \citep{manus2025context,langchain2025context}, independent of the correctness of any single answer. Our experiments isolate one read-only question at a time, a setting in which that benefit cannot express itself as accuracy. We therefore read our result narrowly: on single-question, read-only question answering over an indexable repository, the accuracy cost of the coordination outweighs its context-hygiene benefit. Whether the balance shifts in a long, multi-task session is a question our design cannot answer, and we regard it as an important open one.

A second boundary condition concerns whether an index exists at all. Semantic search presupposes that the repository can be indexed in advance. Agents that operate from a terminal often work over a filesystem that is too large, too volatile, or too arbitrary to index ahead of time, and in that regime dynamic exploration is not a preference but a necessity. That boundary is not only about size. A repository changes, and an index is a snapshot of it. Where the working tree moves quickly, on a branch under active development or in a pipeline that builds many commits a day, the index has to be rebuilt or incrementally updated to stay faithful, and an answer served from a stale index is wrong in a way that is difficult to notice. The monetary side of that upkeep is minor, since only the chunks a change touches need re-embedding and embedding all fifteen repositories in full cost under one dollar in total, but the operational side is real, because change detection, invalidation, and re-embedding are infrastructure that has to be built, run, and monitored. Live exploration carries none of it. It reads whatever is on disk at the moment it is asked, so it is never stale and needs no warm-up, and that is a genuine advantage of the exploration paradigm that neither our accuracy nor our cost figures capture. Our measurements were taken on fixed repository snapshots, which is what the benchmark defines and is also the condition most favourable to an index. Deep Agentic search remains the primary option wherever an index is impractical, or would be stale by the time it is queried. Our comparison is therefore about the regime in which both paradigms are genuinely available, namely a stable and indexable codebase queried in a read-only fashion, and it should not be read as a general verdict against exploration.

Taken together, these considerations point to a pragmatic reading rather than a winner. For the common case of read-only question answering over an indexable repository, an index-backed retrieval agent is the stronger and cheaper default, and the heavier orchestration of a deep agent should be justified by needs that a single-answer accuracy metric does not capture, such as context hygiene across long sessions, environments that cannot be indexed, or tasks that write and execute code. A design that follows naturally from our results is a hybrid that retains a cheap semantic index for retrieval and reserves delegated exploration for the questions, repositories, or tasks that genuinely require it. We leave the construction and evaluation of such a router to future work.

\section{Threats to Validity}
\label{sec:ttv}
\paragraph{Construct validity.} Each paradigm is instantiated through one harness. We deliberately chose widely used, general-purpose open-source implementations rather than a harness tuned by a model provider to its own family, so that neither paradigm benefited from provider-specific optimisation. Other harnesses implement delegation differently, and our results should be read as applying to this pairing rather than to every possible implementation of the two paradigms. The two paradigms differ in both their tool space and their orchestration, and these two aspects co-vary by design. We therefore read the comparison as one between two cohesive, real-world paradigm bundles, one standing for index-backed assistants and one for exploration-based assistants, rather than as a single-factor ablation, because that is the level at which the choice is made in practice.

Within that reading, a harness still carries design decisions that are not part of the paradigm it implements, including the wording of its system prompt, how its tool descriptions are phrased, how sub-agent results are summarised before they are returned, and how many steps it permits. These choices are known to affect agent behaviour, and configuration is itself an object of study for deployed coding agents \citep{santos2026decoding,liu2026diveclaudecode}, so a different implementation of the same paradigm could shift the size of the gap we report. We did not run a sensitivity analysis across alternative deep agent harnesses, such as other production agent frameworks \citep{wang2026openhands}, and we do not claim that the magnitude we measure is invariant to that choice. Three things bound what this can explain. The harness is held constant across all four models within a paradigm, so any overhead it imposes is a fixed property of that arm rather than something that varies with the comparison. The direction of the result is the same for all four models and for both vendors, which a quirk of one model's interaction with one prompt would not produce. And the failure analysis locates the dominant Deep mechanism at the planner to sub-agent hand-off, which is the defining structural feature of the paradigm rather than an incidental setting of the harness. We regard a systematic comparison across harnesses as the natural next study rather than as a gap this one silently leaves open.

\paragraph{Internal validity.} Answers were graded by an automatic judge, which raises the risk that the judge rather than the answer drives the result. We addressed this in three ways: the judge is drawn from a different provider and a different model family than any answering model, so it can neither prefer its own outputs nor favour a family it belongs to, and the one comparison that a family preference could in principle distort, between models of different families, is not a comparison this study makes, since every contrast is paired within a single model; we validated it against an independent human panel on a stratified sample and found substantial to almost perfect agreement (Section~\ref{sec:design}, Table~\ref{tab:judgevalidity}); and we report a non-parametric effect size alongside each significance test. Because every comparison is paired within a model and varies only the paradigm, incidental factors such as answer length act on both conditions at once, and in our data the deep agent scored lower rather than higher, so any preference of the judge for longer answers would work against our conclusion rather than for it. Latency was heavy-tailed and sensitive to the execution environment, so we base no conclusion on it.

\paragraph{External validity.} Our evidence comes from a single benchmark, SWE-QA, four language models, and Python repositories, so the findings may not transfer to other benchmarks, other models, or other languages. Covering more models would strengthen the claim, but running the benchmark and, more importantly, coding agent behaviour and failures from execution traces carry a substantial cost in computation and in manual effort, which bounds how much a single study can cover. We therefore report our findings within the boundary of these models, this benchmark, and these harnesses. These repositories are widely used and are likely to be represented in model training data. Our design limits what that can do to the comparison: every comparison is paired within a model on the same question, so both paradigms draw on the same parametric knowledge, and memorisation raises both arms rather than creating a difference between them. As a further check we split the repositories by provenance, since twelve are drawn from SWE-bench and three were added later from SWE-bench-Live to reduce leakage. Semantic led on both subsets for every model, and the pooled advantage was somewhat larger on the three later repositories, 15.9 against 12.7 points, which is the opposite of what an advantage driven by memorisation would predict. We treat this as a descriptive check rather than a test, since the later subset contains only three repositories and all fifteen are long-standing projects whose code may predate the model cutoffs, so we do not claim that contamination is excluded, only that we find no sign of an asymmetric effect. Repository sizes were measured on the current default branch as a proxy for the benchmark snapshot.

\paragraph{Conclusion validity.} We drew on four models and used paired, non-parametric tests, applying Benjamini and Hochberg false discovery rate correction across the family of comparisons and reporting bootstrap confidence intervals for the effect sizes. The direction of the accuracy result was the same for all four models, which makes it unlikely to be an artifact of a single model or of multiple testing. Because questions are nested within repositories rather than fully independent, we repeated the comparison with a bootstrap that resamples whole repositories, and the score difference remained in favour of Semantic for all four models, with every interval excluding zero. Because the Pass rate depends on a fixed cutoff, we also swept that cutoff from 50 to 90 and found that the direction of the difference held for every model at every threshold (Fig.~\ref{fig:threshold}), so the result is not an artifact of the particular threshold used by the benchmark.

\section{Conclusion and Future Work}
\label{sec:conclusion}
We presented an empirical comparison of Semantic search and Deep Agentic search for repository-level code question answering. Across four language models and fifteen Python repositories, we measured accuracy, cost, agent behaviour, and failure modes on the SWE-QA benchmark.

The results were consistent in direction. For every one of the four models, Semantic search matched or exceeded Deep Agentic search in accuracy, and every paired difference remained significant after false discovery rate correction, with a magnitude that ranged from modest for the Gemini 2.5 pair to large for Gemini 3 Flash. Semantic search also consumed fewer input tokens and cost less per correct answer for every model, and the additional effort spent by the deep agent was not associated with higher accuracy. The failure analysis accounts for the gap rather than merely restating it: the deep agent did not remove failures but added a class of coordination failure, absent from a flat retrieval agent and often silent, that its extra machinery did not offset on this task. These findings apply to read-only question answering over repositories that can be indexed in advance, and they should not be extended to multi-turn code editing or to tasks that execute code. Within that setting, they offer a data-driven counterpoint to the current preference for more elaborate Deep Agentic search over index-backed retrieval, at least for read-only question answering over an indexable repository, while leaving room for Deep Agentic search where an index is impractical or where a long, multi-task session places a premium on keeping the orchestrator's context clean.

Several directions follow from this study. A compute-matched baseline that equalises the token budget across the two paradigms would separate the effect of the search strategy from the effect of the compute it consumes. Extending the comparison beyond Python and to a contamination-controlled benchmark would test how far the findings generalise. Finally, our results motivate a hybrid design that keeps a cheap semantic index for routine retrieval and reserves delegated exploration for the questions, repositories, or tasks that genuinely require it, and building and evaluating such a router is the natural next step.

\begin{acknowledgements}
   This work is supported by the Scientific and Technological Research Council of Turkey (TUBITAK). The authors are grateful for the resources and opportunities provided by the council for this research. The study was carried out in collaboration between the R\&D Center of Intellica Business Intelligence and researchers at Yildiz Technical University.
\end{acknowledgements}

\appendix
\sloppy

\section{Agent harnesses and tool contracts}
\label{app:prompts}

\subsection{Semantic search (ReAct agent)}
The Semantic search agent uses the ReAct agent harness from LangChain (\texttt{create\_agent()}) in the LangChain (version 0.3 and later) and LangGraph ecosystem.
\begin{itemize}
   \item \textbf{Library reference.} LangChain standard agent implementation (\texttt{langchain.agents.create\_agent}).
   \item \textbf{Tool contract.} Restricted strictly to the three tools defined in the SWE-QA benchmark: \texttt{get\_repo\_structure}, \texttt{search\_rag}, and \texttt{read\_file}. The agent has no shell access, no filesystem writing, and no planning tools.
\end{itemize}

\subsection{Deep Agentic search (Deep Agent)}
The Deep Agentic search architecture uses the Deep Agent harness from the official \texttt{langchain-deep-agents} library.
\begin{itemize}
   \item \textbf{Library reference.} Official \texttt{langchain-deep-agents} repository (\url{https://github.com/langchain-ai/deep-agents}).
   \item \textbf{Tool contract.} The default toolset of the framework, comprising task-list planning (\texttt{write\_todos}), sub-agent delegation (\texttt{task}), text search (\texttt{grep}), file listing and globbing (\texttt{ls}, \texttt{glob}), and file reading (\texttt{read\_file}). The orchestrator and the sub-agent are instantiated with the same model.
\end{itemize}

\section{RAG ingestion and retrieval configuration}
\label{app:rag}

\subsection{Chunking strategy}
\begin{itemize}
   \item \textbf{Splitter.} \texttt{RecursiveCharacterTextSplitter} configured for source code.
   \item \textbf{Chunk size.} 500 tokens (about 2{,}000 characters).
   \item \textbf{Chunk overlap.} 50 tokens (about 200 characters), to preserve context continuity across chunk boundaries.
\end{itemize}

\subsection{Embedding model}
\begin{itemize}
   \item \textbf{Model.} Dense vector embeddings using \texttt{text-embedding-3-small} (1{,}536 dimensions).
   \item \textbf{Normalization.} L2 normalized vectors for cosine distance.
\end{itemize}

\subsection{Indexing and vector database}
\begin{itemize}
   \item \textbf{Database engine.} Chroma vector database, instantiated separately per repository.
   \item \textbf{Retrieval metric.} Cosine similarity.
   \item \textbf{Top-$k$ chunks.} $k = 10$ chunks retrieved per \texttt{search\_rag} call.
   \item \textbf{Scope.} Each repository is indexed prior to benchmark execution, so the ingestion cost is paid once per repository.
\end{itemize}

\section{Model parameters and execution environment}
\label{app:env}

\subsection{Inference hyperparameters}
To ensure comparability across all conditions, deterministic sampling parameters were enforced.
\begin{itemize}
   \item \textbf{Temperature.} 0.0 (greedy decoding) across all four answering models (Gemini 2.5 Flash, Gemini 2.5 Pro, Gemini 3 Flash, and Qwen3-235B) and the judge.
   \item \textbf{Top-$p$ and top-$k$.} 1.0 and default.
   \item \textbf{Presence and frequency penalty.} 0.0.
   \item \textbf{Termination.} Each question was bounded by a wall-clock limit of 900 seconds and a recursion limit of 80 steps on the agent graph. At roughly ten seconds for a tool call and its observation, the two bounds are of the same order, so neither is redundant. A run reaching either bound was kept in the results as a failure rather than retried.
\end{itemize}

\subsection{API providers and endpoints}
\begin{itemize}
   \item \textbf{Gemini models.} Gemini 2.5 Flash, Gemini 2.5 Pro, and Gemini 3 Flash are served through Google Cloud Vertex AI using the \texttt{google-genai} and \texttt{langchain-google-genai} SDKs.
   \item \textbf{Qwen3-235B.} Served through an open-weight cloud inference provider (Azure AI Foundry endpoints).
   \item \textbf{Judge.} Claude Sonnet 4.6, served through the Anthropic API. The judge therefore runs on a different provider from every answering model as well as being a different model.
\end{itemize}

\subsection{Execution environment and software dependencies}
\begin{itemize}
   \item \textbf{Python environment.} Python 3.11 or later, managed with \texttt{uv}.
   \item \textbf{Core libraries.} \texttt{langchain} ($\geq$0.3.0), \texttt{langgraph} ($\geq$0.2.0), \texttt{langchain-deep-agents}, and \texttt{pydantic} ($\geq$2.0).
   \item \textbf{Operating system.} Ubuntu Linux 22.04 LTS (x86\_64).
   \item \textbf{Filesystem isolation.} Every question is executed in an isolated thread with a clean workspace containing only the target repository. Cross-repository access is blocked.
\end{itemize}

\section{Judge evaluation and rubric}
\label{app:judge}

\subsection{Judge prompt template}
The judge model is Claude Sonnet 4.6, which is independent of the four answering models and of their providers. The evaluation prompt is reproduced below.

\begingroup\footnotesize
\begin{verbatim}
You are an expert software engineering evaluator. Your task
is to grade a candidate LLM agent's answer to a code question
about a repository against the reference ground-truth answer.

[QUESTION]
{question}

[REFERENCE ANSWER]
{reference_answer}

[CANDIDATE ANSWER]
{candidate_answer}

Evaluate the candidate answer across 5 dimensions on a 0-10
integer scale:
1. Correctness: Does the answer correctly state technical
   facts and mechanisms?
2. Completeness: Does the answer address all parts of the
   question?
3. Relevance: Is the content directly focused on the question
   without fluff?
4. Clarity: Is the explanation well-structured and easy to
   follow?
5. Reasoning: Is the step-by-step logic sound and supported
   by code evidence?

Output your evaluation in valid JSON format:
{
  "correctness": <0-10>,
  "completeness": <0-10>,
  "relevance": <0-10>,
  "clarity": <0-10>,
  "reasoning": <0-10>,
  "rationale": "<brief explanation>"
}
\end{verbatim}
\endgroup

\subsection{Score harmonization and verdict}
The five dimension scores are combined into a total on a 0 to 100 scale:
\[
   \mathrm{Total} = 2 \times (\mathrm{Correctness} + \mathrm{Completeness} + \mathrm{Relevance} + \mathrm{Clarity} + \mathrm{Reasoning}).
\]
The verdict follows the benchmark thresholds:
\begin{itemize}
   \item \textbf{Pass.} Total score $\geq 70$.
   \item \textbf{Partial.} $50 \leq$ Total score $< 70$.
   \item \textbf{Fail.} Total score $< 50$.
\end{itemize}

\section{Cost of running the study}
\label{app:studycost}
Table~\ref{tab:studycost} reports what it cost to produce the measurements in this paper, covering agent inference for all eight conditions and the one-time construction of the vector indexes. All figures are estimates derived from measured token counts and published prices. Agent inference dominates. Index construction is negligible by comparison: embedding all fifteen repositories, about 21 million tokens in total, costs under \$1, and that cost is paid once per repository and then amortised over every question and every condition. Charging the entire index cost to a single Semantic condition would add about \$0.0006 per question, which does not change any of the cost comparisons reported in Section~\ref{sec:results}.

\begin{table}[htbp]
   \centering
   \small
   \caption{Estimated cost of the study in USD, covering agent inference and index construction. Per-question figures are means over the questions in that condition. All values are estimates derived from measured token counts and published prices.}
   \label{tab:studycost}
   \begin{tabular}{llrr}
      \toprule
      Model                                                                     & Paradigm  & Per question & Total    \\
      \midrule
      Gemini 2.5 Flash                                                          & Semantic  & \$0.103      & \$74.09  \\
      Gemini 2.5 Flash                                                          & Deep      & \$0.187      & \$134.53 \\
      Gemini 2.5 Pro                                                            & Semantic  & \$0.568      & \$408.58 \\
      Gemini 2.5 Pro                                                            & Deep      & \$0.675      & \$485.26 \\
      Gemini 3 Flash                                                            & Semantic  & \$0.160      & \$115.24 \\
      Gemini 3 Flash                                                            & Deep      & \$0.341      & \$245.32 \\
      Qwen3-235B                                                                & Semantic  & \$0.009      & \$6.12   \\
      Qwen3-235B                                                                & Deep      & \$0.169      & \$121.92 \\
      \midrule
      \multicolumn{3}{l}{Agent inference, all eight conditions}                 & \$1591.06                           \\
      \multicolumn{3}{l}{Vector index construction, 15 repositories (one time)} & \$0.43                              \\
      \midrule
      \multicolumn{3}{l}{Total}                                                 & \$1591.49                           \\
      \bottomrule
   \end{tabular}
\end{table}

\clearpage

\bibliographystyle{spbasic}
\bibliography{references}

@inproceedings{peng2026swe,
    title = "{SWE}-{QA}: Can Language Models Answer Repository-level Code Questions?",
    author = "Peng, Weihan  and
      Shi, Yuling  and
      Wang, Yuhang  and
      Zhang, Xinyun  and
      Shen, Beijun  and
      Gu, Xiaodong",
    editor = "Liakata, Maria  and
      Moreira, Viviane P.  and
      Zhang, Jiajun  and
      Jurgens, David",
    booktitle = "Findings of the {A}ssociation for {C}omputational {L}inguistics: {ACL} 2026",
    month = jul,
    year = "2026",
    address = "San Diego, California, United States",
    publisher = "Association for Computational Linguistics",
    url = "https://aclanthology.org/2026.findings-acl.402/",
    doi = "10.18653/v1/2026.findings-acl.402",
    pages = "8230--8245",
    ISBN = "979-8-89176-395-1",
}

@unpublished{peng2026deeprepoqa,
  author    = {Peng, Wang and others},
  title     = {{DeepRepoQA}: Code Agent Exploration and Plan-Search via Monte Carlo Tree Search},
  note      = {Under review at ICLR 2026},
  year      = {2026}
}

@article{landis1977,
  author  = {Landis, J. Richard and Koch, Gary G.},
  title   = {The Measurement of Observer Agreement for Categorical Data},
  journal = {Biometrics},
  volume  = {33},
  number  = {1},
  pages   = {159--174},
  year    = {1977},
  doi     = {10.2307/2529310}
}

@misc{havare2025comprehension,
  author        = {Jayant Havare and Saurav Chaudhary and Ganesh Ramakrishnan and Kaushik Maharajan and Srikanth Tamilselvam},
  title         = {A Code Comprehension Benchmark for Large Language Models for Code},
  year          = {2025},
  eprint        = {2507.10641},
  archivePrefix = {arXiv},
  url           = {https://arxiv.org/abs/2507.10641}
}

@misc{ugare2026agentic,
  author        = {Shubham Ugare and Satish Chandra},
  title         = {Agentic Code Reasoning},
  year          = {2026},
  eprint        = {2603.01896},
  archivePrefix = {arXiv},
  url           = {https://arxiv.org/abs/2603.01896}
}

@misc{huang2026atlas,
  author        = {Deli Huang and Cunguang Wang and Hongyin Tang and Zhe Tang and Linsen Guo and Dongyu Ru and Ruoshi Yuan and Ziyue Zhu and Xiaoyu Li and Ziwen Wang and Chen Zhang and Anchun Gui and Wen Zan and Jiaqi Zhang and Xuezhi Cao and Jingang Wang and Xunliang Cai and Yixin Cao},
  title         = {{ATLAS}: All-round Testing of Long-context Abilities across Scales},
  year          = {2026},
  eprint        = {2605.28079},
  archivePrefix = {arXiv},
  url           = {https://arxiv.org/abs/2605.28079}
}

@misc{alebachew2026beyond,
  author        = {Yoseph Berhanu Alebachew and Hunter Leary and Swanand Vaishampayan and Chris Brown},
  title         = {Beyond Code Snippets: Benchmarking {LLMs} on Repository-Level Question Answering},
  year          = {2026},
  eprint        = {2603.26567},
  archivePrefix = {arXiv},
  url           = {https://arxiv.org/abs/2603.26567}
}

@misc{wang2025corag,
  author        = {Liang Wang and Haonan Chen and Nan Yang and Xiaolong Huang and Zhicheng Dou and Furu Wei},
  title         = {Chain-of-Retrieval Augmented Generation},
  year          = {2025},
  eprint        = {2501.14342},
  archivePrefix = {arXiv},
  url           = {https://arxiv.org/abs/2501.14342}
}

@misc{anon2025codeinsightbench,
  author = {{Anonymous}},
  title  = {{CodeInsightBench}: A Benchmark for Advanced Code Understanding and Comparison in Large Language Models},
  year   = {2025},
  note   = {OpenReview submission, under double-blind review},
  url    = {https://openreview.net/forum?id=ThNHBP1qk9}
}

@misc{shi2026codeocr,
  author        = {Yuling Shi and Chaoxiang Xie and Zhensu Sun and Yeheng Chen and Chenxu Zhang and Longfei Yun and Chengcheng Wan and Hongyu Zhang and David Lo and Xiaodong Gu},
  title         = {{CodeOCR}: On the Effectiveness of Vision Language Models in Code Understanding},
  year          = {2026},
  eprint        = {2602.01785},
  archivePrefix = {arXiv},
  url           = {https://arxiv.org/abs/2602.01785}
}

@misc{zhang2026codeqabench,
  author        = {Jun Zhang and JianYing Qu and Hanwen Du and Zhongkai Sun and Yehua Yang and Qiao Zhao},
  title         = {{Code-QA-Bench}: Separating Code Reasoning from Documentation Memorization in Repository-Level {QA}},
  year          = {2026},
  eprint        = {2605.29277},
  archivePrefix = {arXiv},
  url           = {https://arxiv.org/abs/2605.29277}
}

@inproceedings{ahmed2024codeqa,
  author    = {M. Ahmed and M. Dorrah and A. Ashraf and Y. Adel and A. Elatrozy and B. E. Mohamed and W. Gomaa},
  title     = {{CodeQA}: Advanced Programming Question-Answering Using {LLM} Agent and {RAG}},
  booktitle = {2024 6th Novel Intelligent and Leading Emerging Sciences Conference (NILES)},
  pages     = {494--499},
  year      = {2024},
  doi       = {10.1109/NILES63360.2024.10753267}
}

@inproceedings{hu2025coderepoqa,
  author        = {Ruida Hu and Chao Peng and Jingyi Ren and Bo Jiang and Xiangxin Meng and Qinyun Wu and Pengfei Gao and Xinchen Wang and Cuiyun Gao},
  title         = {{CodeRepoQA}: A Large-scale Benchmark for Software Engineering Question Answering},
  booktitle     = {Companion Proceedings of the ACM Web Conference 2025 (WWW '25 Companion)},
  year          = {2025},
  eprint        = {2412.14764},
  archivePrefix = {arXiv},
  url           = {https://arxiv.org/abs/2412.14764}
}

@misc{yang2025codesimpleqa,
  author        = {Jian Yang and Wei Zhang and Yizhi Li and Shawn Guo and Haowen Wang and Aishan Liu and Ge Zhang and Zili Wang and Zhoujun Li and Xianglong Liu and Weifeng Lv},
  title         = {{CodeSimpleQA}: Scaling Factuality in Code Large Language Models},
  year          = {2025},
  eprint        = {2512.19424},
  archivePrefix = {arXiv},
  url           = {https://arxiv.org/abs/2512.19424}
}

@inproceedings{li2025coir,
  author    = {Xiangyang Li and Kuicai Dong and Yi Quan Lee and Wei Xia and Hao Zhang and Xinyi Dai and Yasheng Wang and Ruiming Tang},
  title     = {{CoIR}: A Comprehensive Benchmark for Code Information Retrieval Models},
  booktitle = {Proceedings of the 63rd Annual Meeting of the Association for Computational Linguistics (Volume 1: Long Papers)},
  pages     = {22074--22091},
  year      = {2025},
  doi       = {10.18653/v1/2025.acl-long.1072}
}

@inproceedings{gao2026contextpilot,
  author       = {Shuzheng Gao and Chaozheng Wang and Shuqing Li and Yun Peng and Michael R. Lyu},
  title        = {{ContextPilot}: Code Context Engineering with Memory-Augmented Exploration Agents},
  booktitle    = {Proceedings of the Third International Workshop on Large Language Models for Code (LLM4Code '26)},
  year         = {2026}
}

@misc{wang2026pruning,
  author        = {Jingjing Wang and Xiwen Chen and Wenhui Zhu and Huayu Li and Zhengxiao He and Feiyang Cai and Ana S. Carreon-Rascon and Xuanzhao Dong and Feng Luo},
  title         = {Context Pruning for Coding Agents via Multi-Rubric Latent Reasoning},
  year          = {2026},
  eprint        = {2605.15315},
  archivePrefix = {arXiv},
  url           = {https://arxiv.org/abs/2605.15315}
}

@inproceedings{choi2025corac,
  author    = {YunSeok Choi and CheolWon Na and Jee-Hyong Lee},
  title     = {{CoRAC}: Integrating Selective {API} Document Retrieval with Question Semantic Intent for Code Question Answering},
  booktitle = {Proceedings of the 2025 Conference of the Nations of the Americas Chapter of the Association for Computational Linguistics: Human Language Technologies (Volume 1: Long Papers)},
  pages     = {12620--12635},
  year      = {2025},
  doi       = {10.18653/v1/2025.naacl-long.628}
}

@misc{chen2025coreqa,
  author        = {Jialiang Chen and Kaifa Zhao and Jie Liu and Chao Peng and Jierui Liu and Hang Zhu and Pengfei Gao and Ping Yang and Shuiguang Deng},
  title         = {{CoReQA}: Uncovering Potentials of Language Models in Code Repository Question Answering},
  year          = {2025},
  eprint        = {2501.03447},
  archivePrefix = {arXiv},
  url           = {https://arxiv.org/abs/2501.03447}
}

@article{gong2025cosqa,
  author        = {Jing Gong and Yanghui Wu and Linxi Liang and Yanlin Wang and Jiachi Chen and Mingwei Liu and Zibin Zheng},
  title         = {{CoSQA+}: Enhancing Code Search Evaluation With a Multi-Choice Benchmark and Test-Driven Agents},
  journal       = {IEEE Transactions on Software Engineering},
  year          = {2025},
  eprint        = {2406.11589},
  archivePrefix = {arXiv},
  url           = {https://arxiv.org/abs/2406.11589}
}

@inproceedings{li2026deepagent,
  author        = {Xiaoxi Li and Wenxiang Jiao and Jiarui Jin and Guanting Dong and Jiajie Jin and Yinuo Wang and Hao Wang and Yutao Zhu and Ji-Rong Wen and Yuan Lu and Zhicheng Dou},
  title         = {{DeepAgent}: A General Reasoning Agent with Scalable Toolsets},
  booktitle     = {Proceedings of the ACM Web Conference 2026 (WWW '26)},
  year          = {2026},
  eprint        = {2510.21618},
  archivePrefix = {arXiv},
  url           = {https://arxiv.org/abs/2510.21618}
}

@misc{li2025reposearch,
  author        = {Guochang Li and Yuchen Liu and Zhen Qin and Yunkun Wang and Jianping Zhong and Chen Zhi and Binhua Li and Fei Huang and Yongbin Li and Shuiguang Deng},
  title         = {Empowering {RepoQA}-Agent based on Reinforcement Learning Driven by Monte-Carlo Tree Search},
  year          = {2025},
  eprint        = {2510.26287},
  archivePrefix = {arXiv},
  url           = {https://arxiv.org/abs/2510.26287}
}

@misc{wang2026swdbench,
  author        = {Xinchen Wang and Ruida Hu and Cuiyun Gao and Pengfei Gao and Chao Peng},
  title         = {Evaluating Repository-level Software Documentation via Question Answering and Feature-Driven Development},
  year          = {2026},
  eprint        = {2604.06793},
  archivePrefix = {arXiv},
  url           = {https://arxiv.org/abs/2604.06793}
}

@inproceedings{lopes2025t5,
  author    = {Eduardo Dos S. Lopes and Hil{\'a}rio T. A. de Oliveira and Kelly Assis de Souza Gazolli},
  title     = {Exploring {T5}-Based Code {QA} Systems to Support Teaching Programming in Portuguese and English},
  booktitle = {2025 LI Latin American Computer Conference (CLEI)},
  year      = {2025},
  doi       = {10.1109/CLEI67442.2025.11420747}
}

@misc{li2026fastcode,
  author        = {Zhonghang Li and Zongwei Li and Yuxuan Chen and Han Shi and Jiawei Li and Jierun Chen and Haoli Bai and Chao Huang},
  title         = {{FastCode}: Fast and Cost-Efficient Code Understanding and Reasoning},
  year          = {2026},
  eprint        = {2603.01012},
  archivePrefix = {arXiv},
  url           = {https://arxiv.org/abs/2603.01012}
}

@misc{zhang2026fastcontext,
  author        = {Shaoqiu Zhang and Maoquan Wang and Yuling Shi and Yuhang Wang and Xiaodong Gu and Yongqiang Yao and Tori Gong and Sheng Chen and Rao Fu and Anisha Agarwal and Spandan Garg and Gabriel Ryan and Colin Merkel and Yufan Huang and Shengyu Fu},
  title         = {{FastContext}: Training Efficient Repository Explorer for Coding Agents},
  year          = {2026},
  eprint        = {2606.14066},
  archivePrefix = {arXiv},
  url           = {https://arxiv.org/abs/2606.14066}
}

@misc{sen2026grep,
  author       = {Sahil Sen and Akhil Kasturi and Elias Lumer and Anmol Gulati and Vamse Kumar Subbiah},
  title        = {Is {Grep} All You Need? How Agent Harnesses Reshape Agentic Search},
  year         = {2026},
  eprint       = {2605.15184},
  archivePrefix = {arXiv},
  url          = {https://arxiv.org/abs/2605.15184}
}

@misc{hu2026larger,
  author       = {Yuntong Hu and Tongli Su and Liang Zhao and Bowen Zhu and Hasibul Haque},
  title        = {{LARGER}: Lexically Anchored Repository Graph Exploration and Retrieval},
  year         = {2026},
  eprint       = {2605.16352},
  archivePrefix = {arXiv},
  url          = {https://arxiv.org/abs/2605.16352}
}

@inproceedings{andryushchenko2024codeqa,
  author       = {Georgy Andryushchenko and Vladimir Ivanov and Vladimir Makharev and Elizaveta Tukhtina and Aidar Valeev},
  title        = {Leveraging Large Language Models in Code Question Answering: Baselines and Issues},
  booktitle    = {Analysis of Images, Social Networks and Texts (AIST 2024)},
  publisher    = {Springer},
  year         = {2024},
  doi          = {10.1007/978-3-031-97019-1_1}
}

@misc{usai2026logiclens,
  author       = {Niko Usai and Dario Montagnini and Kristian Ilianov Iliev and Raffaele Camanzo},
  title        = {{LogicLens}: Leveraging Semantic Code Graph to explore Multi Repository large systems},
  year         = {2026},
  eprint       = {2601.10773},
  archivePrefix = {arXiv},
  url          = {https://arxiv.org/abs/2601.10773}
}

@misc{rando2025longcodebench,
  author       = {Stefano Rando and Luca Romani and Alessio Sampieri and Yuta Kyuragi and Luca Franco and Fabio Galasso and Tatsunori Hashimoto and John Yang},
  title        = {{LongCodeBench}: Evaluating Coding {LLMs} at 1M Context Windows},
  year         = {2025},
  eprint       = {2505.07897},
  archivePrefix = {arXiv},
  url          = {https://arxiv.org/abs/2505.07897}
}

@inproceedings{shi2025longcodezip,
  author       = {Yuling Shi and Yichun Qian and Hongyu Zhang and Beijun Shen and Xiaodong Gu},
  title        = {{LongCodeZip}: Compress Long Context for Code Language Models},
  booktitle    = {Proceedings of the IEEE/ACM International Conference on Automated Software Engineering (ASE)},
  year         = {2025},
  doi          = {10.1109/ASE63991.2025.00020}
}

@misc{feng2026compression,
  author       = {Jiazhan Feng and Zhan Qin and Cuiyun Gao and Ruiqi Wang and Chaozheng Wang and Yingwei Ma and Xiaoyuan Xie},
  title        = {On the Effectiveness of Context Compression for Repository-Level Tasks: An Empirical Investigation},
  year         = {2026},
  eprint       = {2604.13725},
  archivePrefix = {arXiv},
  url          = {https://arxiv.org/abs/2604.13725}
}

@misc{shah2025ranger,
  author       = {Pratik Shah and Rajat Ghosh and Aryan Singhal and Debojyoti Dutta},
  title        = {{RANGER}: Repository-Level Agent for Graph-Enhanced Retrieval},
  year         = {2025},
  eprint       = {2509.25257},
  archivePrefix = {arXiv},
  url          = {https://arxiv.org/abs/2509.25257}
}

@inproceedings{yao2023react,
  author       = {Shunyu Yao and Jeffrey Zhao and Dian Yu and Nan Du and Izhak Shafran and Karthik Narasimhan and Yuan Cao},
  title        = {{ReAct}: Synergizing Reasoning and Acting in Language Models},
  booktitle    = {International Conference on Learning Representations (ICLR)},
  year         = {2023},
  eprint       = {2210.03629},
  archivePrefix = {arXiv},
  url          = {https://arxiv.org/abs/2210.03629}
}

@misc{rahman2025refactorcoderqa,
  author       = {Shadikur Rahman and Aroosa Hameed and Gautam Srivastava and Syed Muhammad Danish},
  title        = {{RefactorCoderQA}: Benchmarking {LLMs} for Multi-Domain Coding Question Solutions in Cloud and Edge Deployment},
  year         = {2025},
  eprint       = {2509.10436},
  archivePrefix = {arXiv},
  url          = {https://arxiv.org/abs/2509.10436}
}

@inproceedings{kocadere2026repowise,
  author       = {Mehmet Oguz Kocadere and Yusuf Emir Comert and Sudenaz Yazici and Samil Giray Karacay and Ibrahim Baran Yildiz and Tugba Gurgen Erdogan},
  title        = {{RepoWise}: A Hybrid {RAG}-Based Conversational System for Software Repository Analysis},
  booktitle    = {2026 8th International Congress on Human-Computer Interaction, Optimization and Robotic Applications (ICHORA)},
  publisher    = {IEEE},
  year         = {2026},
  doi          = {10.1109/ICHORA69329.2026.11537220}
}

@misc{gao2023rag,
  author       = {Yunfan Gao and Yun Xiong and Xinyu Gao and Kangxiang Jia and Jinliu Pan and Yuxi Bi and Yi Dai and Jiawei Sun and Meng Wang and Haofen Wang},
  title        = {Retrieval-Augmented Generation for Large Language Models: A Survey},
  year         = {2023},
  eprint       = {2312.10997},
  archivePrefix = {arXiv},
  url          = {https://arxiv.org/abs/2312.10997}
}

@inproceedings{lewis2020retrieval,
  author        = {Lewis, Patrick and Perez, Ethan and Piktus, Aleksandra and Petroni, Fabio and Karpukhin, Vladimir and Goyal, Naman and K{\"u}ttler, Heinrich and Lewis, Mike and Yih, Wen-tau and Rockt{\"a}schel, Tim and Riedel, Sebastian and Kiela, Douwe},
  title         = {Retrieval-Augmented Generation for Knowledge-Intensive {NLP} Tasks},
  booktitle     = {Advances in Neural Information Processing Systems (NeurIPS)},
  year          = {2020},
  eprint        = {2005.11401},
  archivePrefix = {arXiv},
  url           = {https://arxiv.org/abs/2005.11401}
}

@misc{maharaj2026robustness,
  author        = {Kishan Maharaj and Nandakishore Menon and Ashita Saxena and Srikanth Tamilselvam},
  title         = {Robustness and Reasoning Fidelity of Large Language Models in Long-Context Code Question Answering},
  year          = {2026},
  eprint        = {2602.17183},
  archivePrefix = {arXiv},
  url           = {https://arxiv.org/abs/2602.17183}
}

@inproceedings{mohammed2026rubberduck,
  author        = {Ferida Mohammed and Fatma Ayad and Petros Maniatis and Satish Chandra and Elizabeth Dinella},
  title         = {{RubberDuckBench}: A Benchmark for {AI} Coding Assistants},
  booktitle     = {Proceedings of the 3rd International Workshop on Large Language Models for Code (LLM4Code '26)},
  year          = {2026},
  eprint        = {2601.16456},
  archivePrefix = {arXiv},
  url           = {https://arxiv.org/abs/2601.16456}
}

@inproceedings{zhang2026simpledevqa,
  author        = {Jing Zhang and Lianghong Guo and Yanlin Wang and Mingwei Liu and Jiachi Chen and Yuchi Ma and Ensheng Shi and Terry Yue Zhuo and Hongyu Zhang and Zibin Zheng},
  title         = {{SimpleDevQA}: Benchmarking Large Language Models on Development Knowledge {QA}},
  booktitle     = {Findings of the Association for Computational Linguistics: ACL 2026},
  year          = {2026},
  eprint        = {2512.08867},
  archivePrefix = {arXiv},
  url           = {https://arxiv.org/abs/2512.08867}
}

@misc{raghavendra2026sweatlas,
  author        = {Mohit Raghavendra and Soham Dan and Miguel Romero Calvo and Yannis Yiming He and Johannes Baptist Mols and Gautam Anand and Cole McCollum and Edgar Arakelyan and Vijay Bharadwaj and Andrew Park and Jeff Da and MohammadHossein Rezaei and Bing Liu and Brad Kenstler and Yunzhong He},
  title         = {{SWE} Atlas: Benchmarking Coding Agents Beyond Issue Resolution},
  year          = {2026},
  eprint        = {2605.08366},
  archivePrefix = {arXiv},
  url           = {https://arxiv.org/abs/2605.08366}
}

@misc{zhang2026sweexplore,
  author        = {Shaoqiu Zhang and Yuhang Wang and Jialiang Liang and Yuling Shi and Wenhao Zeng and Maoquan Wang and Shilin He and Ningyuan Xu and Siyu Ye and Kai Cai and Xiaodong Gu},
  title         = {{SWE}-Explore: Benchmarking How Coding Agents Explore Repositories},
  year          = {2026},
  eprint        = {2606.07297},
  archivePrefix = {arXiv},
  url           = {https://arxiv.org/abs/2606.07297}
}

@misc{wang2026swepruner,
  author        = {Yuhang Wang and Yuling Shi and Mo Yang and Rongrui Zhang and Shilin He and Heng Lian and Yuting Chen and Siyu Ye and Kai Cai and Xiaodong Gu},
  title         = {{SWE}-Pruner: Self-Adaptive Context Pruning for Coding Agents},
  year          = {2026},
  eprint        = {2601.16746},
  archivePrefix = {arXiv},
  url           = {https://arxiv.org/abs/2601.16746}
}

@misc{elkoussy2026sweqa,
  author        = {La{\"i}la Elkoussy and Julien Perez},
  title         = {{SWE}-{QA}: A Dataset and Benchmark for Complex Code Understanding},
  year          = {2026},
  eprint        = {2604.24814},
  archivePrefix = {arXiv},
  url           = {https://arxiv.org/abs/2604.24814}
}

@inproceedings{cai2026sweqapro,
  author        = {Songcheng Cai and Zhiheng Lyu and Yuansheng Ni and Xiangchao Chen and Baichuan Zhou and Shenzhe Zhu and Yi Lu and Haozhe Wang and Chi Ruan and Benjamin Schneider and Weixu Zhang and Xiang Li and Andy Zheng and Yuyu Zhang and Ping Nie and Wenhu Chen},
  title         = {{SWE}-{QA}-Pro: A Representative Benchmark and Scalable Training Recipe for Repository-Level Code Understanding},
  booktitle     = {Annual Meeting of the Association for Computational Linguistics (ACL 2026)},
  year          = {2026},
  eprint        = {2603.16124},
  archivePrefix = {arXiv},
  url           = {https://arxiv.org/abs/2603.16124}
}

@misc{anthropic2025context,
  author       = {{Anthropic}},
  title        = {Effective Context Engineering for {AI} Agents},
  year         = {2025},
  howpublished = {\url{https://www.anthropic.com/engineering/effective-context-engineering-for-ai-agents}},
  note         = {Anthropic Engineering Blog}
}

@misc{manus2025context,
  author       = {{Manus}},
  title        = {Context Engineering for {AI} Agents: Lessons from Building Manus},
  year         = {2025},
  howpublished = {\url{https://manus.im/blog/Context-Engineering-for-AI-Agents-Lessons-from-Building-Manus}},
  note         = {Manus Blog}
}

@misc{cognition2025multiagents,
  author       = {{Cognition}},
  title        = {Don't Build Multi-Agents},
  year         = {2025},
  howpublished = {\url{https://cognition.ai/blog/dont-build-multi-agents}},
  note         = {Cognition Blog}
}

@misc{langchain2025context,
  author       = {{LangChain}},
  title        = {Context Engineering for Agents},
  year         = {2025},
  howpublished = {\url{https://blog.langchain.com/context-engineering-for-agents/}},
  note         = {LangChain Blog}
}

@inproceedings{oskooei2025hierarchical,
  author       = {Oskooei, Amirkia Rafiei and Yukcu, Selcan and Bozoglan, Mehmet Cevheri and Aktas, Mehmet S.},
  title        = {Repository-Level Code Understanding by {LLMs} via Hierarchical Summarization: Improving Code Search and Bug Localization},
  booktitle    = {Computational Science and Its Applications, {ICCSA} 2025 Workshops, Proceedings, Part I},
  pages        = {88--105},
  year         = {2025},
  publisher    = {Springer},
  address      = {Berlin, Heidelberg},
  isbn         = {978-3-031-97575-2},
  doi          = {10.1007/978-3-031-97576-9\_6}
}

@inproceedings{rafiei2026natural,
  author    = {Rafiei Oskooei, Amirkia and Yukcu, S. Selcan and Bozoglan, Mehmet Cevheri and Aktas, Mehmet S.},
  title     = {Natural Language Summarization Enables Multi-Repository Bug Localization by {LLMs} in Microservice Architectures},
  booktitle = {Proceedings of the 3rd International Workshop on Large Language Models for Code (LLM4Code)},
  pages     = {197--205},
  year      = {2026}
}

@inproceedings{oskooei2026manyshot,
  author    = {Oskooei, Amirkia Rafiei and Cosdan, Kaan Baturalp and Isiktas, Husamettin and Aktas, Mehmet S.},
  title     = {When Many-Shot Prompting Fails: An Empirical Study of {LLM} Code Translation},
  booktitle = {Proceedings of the 1st Workshop on Code Translation, Transformation, and Modernization (ReCode '26)},
  pages     = {38--43},
  year      = {2026},
  publisher = {Association for Computing Machinery},
  address   = {New York, NY, USA},
  isbn      = {9798400724114},
  doi       = {10.1145/3786180.3788314}
}

@misc{liu2026diveclaudecode,
  author        = {Liu, Jiacheng and Zhao, Xiaohan and Shang, Xinyi and Shen, Zhiqiang},
  title         = {Dive into {Claude} {Code}: The Design Space of Today's and Future {AI} Agent Systems},
  year          = {2026},
  eprint        = {2604.14228},
  archivePrefix = {arXiv},
  primaryClass  = {cs.SE},
  url           = {https://arxiv.org/abs/2604.14228}
}

@inproceedings{santos2026decoding,
  author    = {dos Santos, H{\'e}lio Victor Flexa and Costa, Vitor and Montandon, Jo{\~a}o Eduardo and Valente, Marco T{\'u}lio},
  title     = {Decoding the Configuration of {AI} Coding Agents: Insights from {Claude} {Code} Projects},
  booktitle = {Proceedings of the 2026 International Workshop on Agentic Engineering (AGENT@ICSE '26)},
  pages     = {63--67},
  year      = {2026},
  publisher = {Association for Computing Machinery},
  address   = {New York, NY, USA},
  doi       = {10.1145/3786167.3788412}
}

@inproceedings{wang2026openhands,
  author    = {Wang, Xingyao and Rosenberg, Simon and Michelini, Juan and Smith, Calvin and Tran, Hoang H. and Nyst, Engel and Malhotra, Rohit and Zhou, Xuhui and Chen, Valerie and Brennan, Robert and Neubig, Graham},
  title     = {The {OpenHands} Software Agent {SDK}: A Composable and Extensible Foundation for Production Agents},
  booktitle = {Proceedings of Machine Learning and Systems 8 (MLSys 2026)},
  year      = {2026},
  url       = {https://proceedings.mlsys.org/paper_files/paper/2026/hash/8ae9cf363ea625161f885b798c1f1f78-Abstract-Conference.html}
}

\end{document}